\documentclass[
 aps,
 jmp,
 amsmath,amssymb,
 twocolumn,
]{revtex4-1}

\usepackage{graphicx}
\usepackage{dcolumn}
\usepackage{bm}
\usepackage{verbatim}
\usepackage{color}

\usepackage{hyperref}
\usepackage{soul}
\usepackage[makeroom]{cancel}

\newcommand{\figref}[1]{FIG.~\ref{#1}}
\newcommand{\refeq}[1]{\eqref{#1}}
\newcommand{\secref}[1]{Section~\ref{#1}}
\newcommand{\appref}[1]{Appendix~\ref{#1}}
\newcommand{\tabref}[1]{TABLE~\ref{#1}}
\newcommand{\figscale}{0.48}

\newcommand{\pdv}[2]{\frac{\partial{#1}}{\partial{#2}}}
\newcommand{\vect}[1]{\boldsymbol{#1}}
\newcommand{\matr}[1]{\mathbf{#1}}
\newcommand{\dI}{\text{d}}
\newcommand{\odv}[2]{\frac{\dI #1}{\dI #2}}
\newcommand{\ddv}[2]{\odv{#1}{#2}}
\newcommand{\mfpe}{\lambda_e}
\newcommand{\mfpei}{\lambda_{ei}}
\newcommand{\Zbar}{Z}
\newcommand{\nue}{\nu_{e}}
\newcommand{\nuei}{\nu_{ei}}
\newcommand{\nuscat}{\nu_{scat}}
\newcommand{\vmag}{v}
\newcommand{\vth}{v_{th}}
\newcommand{\vtwoh}{v_{2 th}}
\newcommand{\vn}{\vect{n}}
\newcommand{\E}{\vect{E}}
\newcommand{\B}{\vect{B}}
\newcommand{\omegaB}{\vect{\omega}_{B}}
\newcommand{\Ez}{E_z}
\newcommand{\qe}{q_e}
\newcommand{\me}{m_e}
\newcommand{\Te}{T_e}
\newcommand{\Ti}{T_i}
\newcommand{\ed}{n_e}
\newcommand{\kB}{k_B}
\newcommand{\fM}{f_M}
\newcommand{\fzero}{f_0}
\newcommand{\vfzero}{\vect{f_0}}
\newcommand{\fone}{{\vect{f_1}}}
\newcommand{\fonez}{f_{1_z}}
\newcommand{\vv}{\vect{v}}
\newcommand{\vvb}{\tilde{\vect{v}}}
\newcommand{\gv}{\nabla_{\vv}}
\newcommand{\gvb}{\nabla_{\vvb}}
\newcommand{\gx}{\nabla_{\vect{x}}}
\newcommand{\ft}{f}
\newcommand{\lnc}{\text{ln}\Lambda}
\newcommand{\Iohm}{\matr{J}_{Ohm}}

\begin{document}

\preprint{AIP/123-QED}

\title[AWBS kinetic modeling of electrons with nonlocal Ohm's law in plasmas relevant to inertial confinement fusion]{AWBS kinetic modeling of electrons with nonlocal Ohm's law in plasmas relevant to inertial confinement fusion}

\author{M. Holec}
 \email{holec1@llnl.gov}
 \affiliation{
Center for Applied Scientific Computing, Lawrence Livermore National
Laboratory, P.O. Box 808, L-561, Livermore, CA 94551, USA.
 }
 \affiliation{
Centre Lasers Intenses et Applications, Universite de Bordeaux-CNRS-CEA,\\ 
UMR 5107, F-33405 Talence, France.
}

\author{P. Loiseau}
 \affiliation{
 CEA, DAM, DIF, F-91297 Arpajon Cedex, France.
}

\author{J. P. Brodrick}
 \affiliation{
York Plasma Institute, Department of Physics, University of York,\\ 
Heslington, York, YO10 5DD, UK.
}

\author{D. Del Sorbo}
 \affiliation{
York Plasma Institute, Department of Physics, University of York,\\ 
Heslington, York, YO10 5DD, UK.
}

\author{A. Debayle}
 \affiliation{
 CEA, DAM, DIF, F-91297 Arpajon Cedex, France.
}

\author{V. Tikhonchuk}
 \affiliation{
Centre Lasers Intenses et Applications, Universite de Bordeaux-CNRS-CEA,\\ 
UMR 5107, F-33405 Talence, France.
}
\affiliation{ 
ELI-Beamlines Institute of Physics, AS CR, v.v.i, 
Na Slovance 2, Praha 8, 180 00, Czech Republic.
}

\author{J.-L. Feugeas}
 \affiliation{
Centre Lasers Intenses et Applications, Universite de Bordeaux-CNRS-CEA,\\ 
UMR 5107, F-33405 Talence, France.
}

\author{Ph. Nicolai}
 \affiliation{
Centre Lasers Intenses et Applications, Universite de Bordeaux-CNRS-CEA,\\ 
UMR 5107, F-33405 Talence, France.
}

\author{B. Dubroca}
\affiliation{
Centre Lasers Intenses et Applications, Universite de Bordeaux-CNRS-CEA,\\ 
UMR 5107, F-33405 Talence, France.
}

\author{C. P. Ridgers}
 \affiliation{
York Plasma Institute, Department of Physics, University of York,\\ 
Heslington, York, YO10 5DD, UK.
}

\author{R. J. Kingham}
 \affiliation{
Plasma Physics Group, Blackett Laboratory, Imperial College,\\ 
London SW7 2BW, United Kingdom.
}

\date{\today}

\begin{abstract}
The~interaction of lasers with plasmas very often leads to nonlocal transport 
conditions, where the classical hydrodynamic model fails to describe 
important microscopic physics related to highly mobile particles. 
In this study we analyze and further propose a~modification of 
the~Albritton-Williams-Bernstein-Swartz collision operator 
Phys. Rev. Lett 57, 1887 (1986)
for the~nonlocal electron transport 
under conditions relevant to ICF. The~electron distribution function
provided by this modification exhibits some very desirable properties 
when compared to the~full Fokker-Planck operator in the~local diffusive regime,
and also performs very well when benchmarked against Vlasov-Fokker-Planck 
and collisional PIC codes in the~nonlocal transport regime, where we find that
the~effect of the~electric field via the~nonlocal Ohm's law is an~essential 
ingredient in order to capture the~electron kinetics properly.
\end{abstract}

\pacs{Valid PACS appear here}

\keywords{kinetics; nonlocal electron transport; laser-heated plasmas; hydrodynamics, Ohm's law.}

\maketitle

\section{Introduction}
\label{sec:Intro}
The~first modern attempts at kinetic modeling of plasma can be traced back 
to the~fifties, when Cohen, Spitzer, and Routly (CSR) \cite{CSR_1950} 
demonstrated that the~effect of Coulomb collisions between electrons and ions 
in the~ionized gas predominantly results 
from frequently occurring events of cumulative small deflections 
rather than occasional close encounters. This effect was originally described
by Jeans in \cite{Jeans_BOOK1929} and 
Chandrasekhar \cite{Chandrasekhar_RMP1943} 
proposed to use the~diffusion equation model of the~Vlasov-Fokker-Planck type 
(VFP)~\cite{Planck_1917}.

A~classical paper by Spitzer and H\"arm (SH) 
\cite{SpitzerHarm_PR1953} provides the~computation of 
the~electron distribution function (EDF) in a~plasma (from low to high $\Zbar$)
with a~temperature gradient accounting for e-e and e-i collisions.
The~resulting expressions for current and heat flux are widely used in plasma 
hydrodynamic models.

The~distribution function based on the spherical harmonics method in 
its first approximation (P1) \cite{Jeans_MNRAS1917} is of the form 
$f^0+\mu f^1$, where $f^0$ and $f^1$ 
are isotropic and $\mu$, is the direction cosine between the particle 
velocity and the~temperature gradient. It should be emphasized that
the~SH solution assumes a~small perturbation of equilibrium, i.e. that 
$f^0$ is the~Maxwell-Boltzmann distribution and $\mu f^1$ represents 
a~very small anisotropic deviation. 
This approximation holds for $L_T\gg\mfpe$, 
a~condition which is often invalid in laser plasmas, 
where $L_T$ is the~temperature length scale and $\mfpe$ 
the~mean free path of electrons. It is worth mentioning, that electrons having
3 to 4 times the~thermal velocity are dominantly responsible for heat-flow
and that those faster than 6 times the~thermal velocity can be completely 
neglected in this local theory.

The~actual cornerstone of the~modern VFP simulations was set in place
by Rosenbluth \cite{Rosenbluth_PR1957}, when he derived a~simplified form 
of the~VFP equation for a~finite expansion of the~distribution function,
where all the~terms are computed according to plasma conditions, including
$f^0$, which of course needs to tend to the~Maxwell-Boltzmann distribution.
Consequently, the~pioneering work on numerical solution of the~VFP equation
\cite{Bell_1981_83, Matte_1982_86} revealed the~importance of the~nonlocal
electron transport in laser-heated plasmas. 
In particular, that the~heat flow down steep temperature gradients in 
unmagnetised plasma cannot be described by the classical, local fluid
description of transport \cite{SpitzerHarm_PR1953, Braginskii_1965_3}.
This is due to the~classical $f^1$ \textit{not being} a~small deviation 
(especially for electrons having 3 to 4 times the~thermal velocity), 
i.e. $f^0\sim f^1$ characterized by $L_T\sim\mfpe$.
It was also shown that a~thermal transport inhibition \cite{Bell_1981_83} 
around the peak of the temperature gradient, and a~nonlocal preheat 
ahead of the main heat wave front, naturally appear. 
These effects are attributed to significant deviations 
of $f^0$ from Maxwellian distribution.

Nevertheless, numerical solution of the~VFP equation even in the~Rosenbluth
formalism remains very challenging computationally, because the~e-e collision
integral is nonlinear. More simple linear forms of e-e collision operator
are needed. Although some VFP simulations on experimentally relevant timescales 
have been performed (for recent examples see 
\cite{Hawreliak04,Ridgers08,Willingale10,Bissell10,Joglekar14,Joglekar16,Henchen_PRL2018}, 
an extensive review has been conducted by Thomas et al. \cite{Thomas13}), 
their relative computational inefficiency severely limits the range of 
simulations that can be performed.

It is the~purpose of this paper to use an~efficient alternative 
to a full solution of the VFP equation introduced in \cite{Sorbo_2015}
to accurately calculate nonlocal transport, 
based on the~Albritton-Williams-Bernstein-Swartz 
collision operator (AWBS) \cite{AWBS_PRL1986}.
In Section \ref{sec:AWBSmodel} we propose a~modified form of 
the~AWBS collision operator. Its important properties are further
presented in Section \ref{sec:DiffusiveKinetics} with the~emphasis on its
comparison to the~full VFP solution in the~local diffusive regime.
In \secref{sec:AWBSnonlocal} we define a~full model of electron kinetics and 
the~way of discretizing the~electron phase-space and also
the~coupling of the~kinetic model to magneto-hydrodynamics.
Section \ref{sec:BenchmarkingAWBS} focuses on the~performance of the~AWBS 
transport equation model compared to modern kinetic codes including VFP codes
Aladin and Impact \cite{Kingham_JCP2004}, and PIC code Calder 
\cite{Perez_PoP2012}, where the~cases related to real
laser generated plasma conditions are studied. Finally, the~most important
outcomes of our research are concluded in Section \ref{sec:Conclusions}. 

\section{The~AWBS kinetic model}
\label{sec:AWBSmodel}

The~electrons in plasma can be modeled by the~deterministic Vlasov model 
of charged particles
\begin{equation}
  \pdv{\ft}{t} + \vv\cdot\gx \ft + 
  \frac{\qe}{\me}\left(\E + \frac{\vv}{c}\vect{\times}\B\right)\cdot\gv \ft = 
  C_{ee}(\ft) + C_{ei}(\ft) ,
  \label{eq:kinetic_equation}
\end{equation}
where $\ft(t, \vect{x}, \vect{v})$ represents 
the~density function of electrons (EDF)
at time $t$, spatial point $\vect{x}$, and velocity $\vv$, $\E$ and $\B$ are 
the~electric and magnetic fields in plasma, $\qe$ and $\me$ being 
the~charge and mass of electron.

The~general form of the~e-e collision operator 
$C_{ee}$ is the~Fokker-Planck form published by Landau \cite{Landau_1936}
\begin{equation}
  C_{FP}(\ft) =
  \Gamma~\gv\cdot\int \matr{U}(\vv - \vvb) \cdot \left(
  \ft\, \gvb \ft - \ft\, \gv \ft \right)\, \dI\vvb ,
  \label{eq:LFP_model}
\end{equation}
where $\Gamma = \frac{4\pi\qe^4\lnc}{\me^2}$, $\lnc$ is the~Coulomb logarithm,
and $\matr{U}(\vv - \vvb) = \frac{1}{|\vv - \vvb|}\left(\matr{I} - \frac{(\vv - \vvb)\otimes (\vv - \vvb)}{|\vv - \vvb|^2}\right)$.
The~e-i collision operator in principle also depends 
on the~ion density function, i.e. $C_{ei}(\ft, \ft_i)$, however it 
can be expressed in a~simpler form independent from $\ft_i$
since massive ions are considered 
to be motionless compared to electrons during a collision. 
The~operator then accounts
for the~change of electron velocity without change in the~velocity magnitude
, i.e. angular scattering. 
It is expressed in spherical coordinates as
\begin{equation}
  C_{ei}(\ft) = \frac{\nuei}{2}
  \left(\pdv{}{\mu}\left((1 - \mu^2)\pdv{\ft}{\mu}\right)
  + \frac{1}{1 - \mu^2}\frac{\partial^2 \ft}{\partial\theta^2} \right) ,
  \label{eq:ei_scattering}
\end{equation}
where $\mu = \cos\phi$, $\phi$ and $\theta$ are the~polar and azimuthal 
angles, and $\nuei = \frac{\Zbar n_e \Gamma}{\vmag^3}$ is the~e-i
collision frequency.

The~e-e collision operator needs to be linearized for efficient computation.
Fisch introduced in \cite{Fisch_RMP1987} a~linear form of 
the~electron-electron collision operator  
in the~high-velocity limit ($\vmag\gg\vth$)
\begin{multline}
  C_{H}(\ft) = \vmag \nue \pdv{}{\vmag}\left(\ft + 
  \frac{\vth^2}{\vmag}\pdv{f}{\vmag}\right) \\
  + \frac{\nue}{2}\left( 1 - \frac{\vth^2}{2\vmag^2}\right) 
  \left(\pdv{}{\mu}\left((1 - \mu^2)\pdv{f}{\mu}\right)
  + \frac{1}{1-\mu^2}\frac{\partial^2f}{\partial\theta^2} \right)
  , \label{eq:HighVelocity_model}
\end{multline}
where $\nue = \frac{n_e \Gamma}{\vmag^3}$ is the~e-e collision 
frequency and $\vth = \sqrt{\frac{\kB T_e}{\me}}$ is the~electron thermal 
velocity and $\kB$ is the~Boltzmann constant.
The~linear form of $C_{H}$ arises from an~assumption that the~fast electrons 
predominantly interact with the~thermal (slow) electrons, 
which is an important simplification to the~form \eqref{eq:LFP_model}.
However the~diffusion term in the~e-e collision operator 
\eqref{eq:HighVelocity_model} still presents numerical difficulties.

A~yet simpler form of the~collision operator of electrons was proposed in 
\cite{Sorbo_2015}
\begin{multline}
  C_{AWBS}(\ft) = \vmag \nue^*\pdv{}{\vmag}\left(\ft - \fM\right) \\
  + \frac{\nuei + \nue^*}{2} 
  \left(\pdv{}{\mu}\left((1 - \mu^2)\pdv{f}{\mu}\right)
  + \frac{1}{1-\mu^2}\frac{\partial^2f}{\partial\theta^2} \right)
  , \label{eq:AWBS_model}
\end{multline}
where $\fM = \frac{n_e}{(2\pi)^{\frac{3}{2}}\vth^3}
\exp\left(-\frac{\vmag^2}{2\vth^2}\right)$ 
is the~Maxwell-Boltzmann equilibrium distribution.
Here, the~first term representing the AWBS operator \cite{AWBS_PRL1986}
accounts for relaxation to equilibrium due to the~e-e collisions, and 
the~second term accounts for the~e-i and e-e collisions contribution 
to scattering.

A~method of angular momenta for the~solution of the~electron kinetic equation
with the~collision operator \eqref{eq:AWBS_model} 
was introduced in \cite{Sorbo_2015, Sorbo_2016}. 

In \eqref{eq:AWBS_model} we have introduced a~modified e-e collision frequency
$\nue^*$ in order to account for a~dependence with respect to 
the~ion charge $\Zbar$ of the~electron thermal conductivity. This issue
is further analyzed in Section \ref{sec:DiffusiveKinetics} and promising 
results compared to the~full FP operator are presented.

\section{BGK, AWBS, and Fokker-Planck models in local diffusive regime}
\label{sec:DiffusiveKinetics}
An~approximate solution to the~\textit{local diffusive regime} 
of electron transport can be found, since it
refers to a~low anisotropy modeled by the~P1 form of EDF  
\begin{equation}
  \tilde{\ft}(z, \vmag, \mu) = \ft^0(z, \vmag) + \mu \ft^1(z, \vmag),
  \label{eq:f_approximation}
\end{equation}
where $z$ is the~spatial coordinate along the~axis $z$, $\vmag$ 
the~magnitude of the~electron velocity. 

The~approximate transport solution is then obtained when analyzing 
the~stationary form of \eqref{eq:kinetic_equation} in one spatial dimenstion
(1D) 
\begin{equation}
  \mu\left(\pdv{\tilde{f}}{z} 
  + \frac{\qe\Ez}{\me\vmag}\pdv{\tilde{f}}{\vmag}\right) 
  + \frac{\qe\Ez}{\me}\frac{(1-\mu^2)}{\vmag^2}\pdv{\tilde{f}}{\mu}
  = \frac{1}{\vmag}C(\tilde{\ft}) ,
  \label{eq:1D_kinetic_equation}
\end{equation}
where $C$ is a~given collision operator including both e-e and e-i collisions.
Condition of plasma \textit{quasi-neutrality}, 
represented by the~zero current 
$\vect{j} \equiv \qe \int \vect{v} \ft \dI\vect{v} = \vect{0}$ 
in the case of an~unmagnetised plasma in 1D according to \eqref{eq:Ampere}, 
is for the~P1 \eqref{eq:f_approximation} expressed as
\begin{equation}
  \int \vmag f^1 \vmag^2 \dI\vmag = 0 ,
  \label{eq:j0_P1}
\end{equation}
and is accounted for by the~effect of $\Ez$ in \eqref{eq:1D_kinetic_equation}.

The~locality of transport is the~best expressed in terms of the~Knudsen number
$\text{Kn}=\frac{\lambda}{L}$, where $\lambda$ is the~mean free path of electron and
$L$ the~characteristic length scale of plasma. Consequently, plasma conditions
characterized by $text{Kn}\ll1$ correspond to a~local transport regime. 
This measure then
play a~very important role in our analysis, where we use the~electron-electron
and electron-ion mean free paths $\mfpe = \Zbar\mfpei = \frac{\vmag}{\nue}$,
and the~density and temperature plasma scale lengths 
$L_{n_e} = n_e/\pdv{n_e}{z}$ and $L_{T_e} = T_e/\pdv{T_e}{z}$.

In practice, the~Knudsen number of thermal electrons is often used as 
a~measure of the~locality of transport corresponding to given plasma conditions,
where $\text{Kn}(\vth)<0.001$ is considered the~limit of validity of 
the~local transport theory \cite{LMV_1983_7}.

\subsection{BGK local diffusive electron transport}
\label{sec:BGKDiffusiveRegime}

Bhatnagar, Gross, and Krook (BGK) introduced a~very simple form
of a~collision operator \cite{BGK_1954}
\begin{equation}
  C_{BGK}(\tilde{\ft})
  =
  \nue(\fM - \tilde{\ft})
  + \frac{\nuei + \nue}{2}
  \pdv{}{\mu}(1 - \mu^2)\pdv{\tilde{\ft}}{\mu} .
  \label{eq:BGK_model_1D}
\end{equation}
In spite of its~simple form, BGK collision operator \eqref{eq:BGK_model_1D} 
serves as a~useful model providing a~relevant kinetic response, yet only 
qualitative with respect to the~FP collision operator \eqref{eq:LFP_model}.
In particular, the~conservation of kinetic energy, momentum, 
and number of particles is often violated 
\cite{Shkarofsky_Particle_Kinetics_book_1966_24}.

However, the~form of \eqref{eq:BGK_model_1D} provides a~simple analytical 
treatment of the~local diffusive transport regime, when used in 
\eqref{eq:1D_kinetic_equation}. As a~result, one finds a~simple form of
the~BGK isotropic and anisotropic terms of \eqref{eq:f_approximation} to be
\begin{eqnarray}
  \ft^0 &=& \fM ,
  \label{eq:BGK_f0} \\
  \ft^1 &=& - \frac{\mfpe}{\Zbar + 2}
  \left( \pdv{\fM}{z} + \frac{\qe\Ez}{\me\vmag}\pdv{\fM}{\vmag} \right) ,
  \label{eq:BGK_f1}
\end{eqnarray}
where a~detailed derivation of \eqref{eq:BGK_f0} and \eqref{eq:BGK_f1} 
can be found in Appendix~\ref{app:DiffusiveKinetics}.
When the~quasi-neutrality constraint \eqref{eq:j0_P1} imposed by $\E_L$
\eqref{app_eq:BGK_Efield} is used, one finally obtains 
the~analytical BGK form of the~anisotropic term
\begin{equation}
  \ft^1 = - \mu 
  \left( \frac{\vmag^2}{2 \vth^2} - 4\right)\frac{1}{\Zbar + 2}
  \frac{\mfpe}{L_{T_e}}\fM
  . 
  \label{eq:BGK_approximation}
\end{equation}
The~details about the~BGK distribution function compared to other
collision operators can be found in Section~\ref{sec:SummaryDiffusiveKinetics}.

\subsection{AWBS local diffusive electron transport}
\label{sec:AWBSDiffusiveRegime}
Similarly to the~BGK model, the~AWBS collision operator \ref{eq:AWBS_model} 
explicitly uses equilibration to the~Maxwell-Boltzmann distribution $\fM$. 
On the~other hand, AWBS originates from $C_H$, which is derived from 
the~full FP operator \eqref{eq:LFP_model}. This makes the~AWBS operator 
to be superior to the~BGK operator, which is considered a~purely
phenomenological model.

If \eqref{eq:AWBS_model} is used in \eqref{eq:1D_kinetic_equation}, one obtains
the~following equations governing the~AWBS isotropic and anisotropic terms of 
\eqref{eq:f_approximation}
\begin{eqnarray}
  \pdv{f^0}{\vmag} &=& \pdv{\fM}{\vmag} ,
  \label{eq:AWBS_f0} \\
  \pdv{f^1}{\vmag}  
  - \frac{\Zbar +  r_A}{\vmag r_A} f^1 &=& 
  \frac{\mfpe}{\vmag r_A}
  \left(\pdv{\fM}{z} + \frac{\qe\Ez}{\me\vmag}\pdv{\fM}{\vmag}\right) ,
  \label{eq:AWBS_f1} 
\end{eqnarray}
where $ r_A$ represents a~scaling parameter defining the~modified
e-e collision frequency as $\nue^* =  r_A \nue$.
A~detailed derivation of \eqref{eq:AWBS_f0} and \eqref{eq:AWBS_f1} 
can be found in Appendix~\ref{app:DiffusiveKinetics}.
Consequently, one finds the~AWBS model equation for $\ft^1$ 
in local diffusive regime to be
\begin{multline}
  \pdv{\ft^1}{\vmag} 
  - \frac{\Zbar +  r_A}{\vmag r_A}\ft^1
  = \\
  \frac{\mfpe}{\vmag r_A}\left(\frac{1}{L_{n_e}} + 
  \left( \frac{\vmag^2}{2 \vth^2} - \frac{3}{2}\right)
  \frac{1}{L_{T_e}} - \frac{\qe\Ez}{\me\vth^2}\right)\fM .
  \label{eq:AWBS_f1_ODE}
\end{multline}
The~solution of \eqref{eq:AWBS_f1_ODE} can be found in terms of 
upper incomplete gamma function $\tilde{\Gamma}$ 
(see \appref{app:DiffusiveKinetics})
\begin{equation}
  \ft^1_{\text{AWBS}} = - \frac{d}{\vmag^a} 
  \left(b~\tilde{\Gamma}{\left(\frac{a+6}{2}, \frac{\vmag^2}{2\vth^2}\right)}
  + c~\tilde{\Gamma}{\left(\frac{a+4}{2}, \frac{\vmag^2}{2\vth^2}\right)}\right)
  ,
  \label{eq:AWBS_analytic_solution}
\end{equation}
where 
$a = -\frac{\Zbar + r_A}{r_A}$, $b = \frac{1}{L_{\Te}}$, 
$c = \frac{1}{L_{\ed}} - \frac{3}{2}\frac{1}{L_{\Te}} 
- \frac{\qe\Ez}{\me\vth^2}$, and
$d = \frac{2^{\frac{a + 2}{2}} \vth^{a + 1}}{r_A (2\pi)^{\frac{3}{2}} \Gamma}$.
Nevertheless, a~numerical solution of \eqref{eq:AWBS_f1_ODE} needs to be 
adopted for higher $\Zbar$  (see \appref{app:DiffusiveKinetics}).
The~\textit{quasi-neutrality} constraint \eqref{eq:j0_P1} applied to
$\ft^1_{\text{AWBS}}$ leads to $\Ez = \E_L$ \eqref{app_eq:BGK_Efield} 
independently from $\Zbar$ and $r_A$. 

\subsection{Fokker-Planck local diffusive electron transport}
\label{sec:FPDiffusiveRegime}

Solution to the~1D transport equation \eqref{eq:1D_kinetic_equation}
using the~Fokker-Planck collision operator \eqref{eq:LFP_model}
is very ambitious, as demonstrated in 
\cite{Chandrasekhar_RMP1943, CSR_1950, Rosenbluth_PR1957}, fortunately, one 
can use the~explicit evaluation of the~electron distribution function
published in \cite{SpitzerHarm_PR1953}, which takes the~following form
\begin{multline}
  \ft^1_{\text{SH}} =
  \frac{\vtwoh^4}{\Gamma\Zbar n_e}\\
  \left( 2\tilde{D}_T\left(\frac{\vmag}{\vtwoh}\right) 
  + \frac{3}{2}\frac{\gamma_T}{\gamma_E} 
  \tilde{D}_E\left(\frac{\vmag}{\vtwoh}\right) \right)
  \frac{\fM}{T}\pdv{T_e}{z}  ,
  \label{eq:f1_SH}
\end{multline}
where $\tilde{D}_T(x) = \Zbar D_{T}(x) / B$, 
$\tilde{D}_E(x) = \Zbar D_{E}(x) / A$, $\gamma_T$,
and $\gamma_E$ are numerical values in TABLE~I, TABLE~II, and
TABLE~III in \cite{SpitzerHarm_PR1953}, and 
$\vtwoh = \sqrt{\frac{\kB T_e}{2\me}}$.

One should be aware, that the~solution of \eqref{eq:1D_kinetic_equation}
with the~full FP collision operator reveals importance of
e-e Coulomb collisions, which is emphasized in the~$\Zbar$ dependence 
of the~distribution function, current, heat flux, 
electric field according to \eqref{eq:j0_P1}, etc.
In particular, the latter exhibits the~following dependence 
\cite{SpitzerHarm_PR1953}
\begin{equation}
  \E = \frac{\me \vth^2}{\qe}
  \left(\frac{\nabla n_e}{n_e} + 
  \left(1 + \frac{3}{2}\frac{\Zbar + 0.477}{\Zbar + 2.15} \right)
  \frac{\nabla T_e}{T_e} \right),
  \label{eq:SH_Efield} 
\end{equation}
which for $\Zbar\gg1$ corresponds to the~classical Lorentz electric field 
\eqref{app_eq:BGK_Efield}.

\subsection{Summary of the~BGK, AWBS, and Fokker-Planck local diffusive 
transport}
\label{sec:SummaryDiffusiveKinetics}

Ever since the~SH paper \cite{SpitzerHarm_PR1953}, the~effect of microscopic
electron transport on the~current $\int \qe \vv \tilde{\ft} \, \dI\vv$ 
and the~heat flux $\int \frac{\me |\vv|^2}{2} \vv \tilde{\ft} \, \dI\vv$ 
in plasmas
under local diffusive conditions has been understood. By overcoming some 
delicate aspects of the~numerical solution to \eqref{eq:LFP_model} presented 
in \cite{CSR_1950}, the~effect of electron-electron collisions
was quantified and dependence on $\Zbar$ of the~heat flux
$\vect{q}$ was approximated as \cite{SpitzerHarm_PR1953, Epperlein_PoFB1991} 
\begin{equation}
  \vect{q} = \xi(\Zbar)~\vect{q}_L 
  = \frac{\Zbar + 0.24}{\Zbar + 4.2} \vect{q}_L ,
  \label{eq:qSH_approximation}
\end{equation}
where 
$\xi$ is the~$\Zbar$-dependence \cite{Epperlein_PoFB1991} approximation and
$\vect{q}_L$
is the~heat flux given 
by the~Lorentz gas model \cite{Lorentz_1905} 
\begin{equation}
  \ft^1_{\text{Lorentz}} = - \mu 
  \left( \frac{\vmag^2}{2 \vth^2} - 4\right)
  \frac{\mfpei}{L_{T_e}}\fM
  . 
  \label{eq:Lorentz_approximation}
\end{equation}
 
In the~case of BGK the~collision operator \eqref{eq:BGK_model_1D} needs to be 
corrected in order to provide a~same local behavior as 
\eqref{eq:qSH_approximation}, i.e. a~correct dependence on 
$\Zbar$. Consequently, we define a~scaling formula 
\begin{equation}
  r_B(\Zbar) = \frac{\zeta\Zbar}{\xi(\Zbar + 2 \zeta)} ,
  \label{eq:BGK_r_scaling}
\end{equation}
based on comparison of the~formula \eqref{eq:BGK_approximation} to 
$\xi \ft^1_{\text{Lorentz}}$ and we write a~consistent local diffusion version
of BGK 
\begin{equation}
  C_{BGK}(\tilde{\ft})
  =
  r_B \nue(\fM - \tilde{\ft})
  + \frac{r_B}{\zeta}\frac{\nuei + \zeta\nue}{2}
  \pdv{}{\mu}(1 - \mu^2)\pdv{\tilde{\ft}}{\mu}
   ,
  \label{eq:BGK_scaling}
\end{equation}
where the~constant $\zeta$ can be set arbitrarily, because it does not affect 
the~local EDF of \eqref{eq:BGK_scaling}
\begin{equation}
  \ft^1_{\text{BGK}} = - \mu 
  \left( \frac{\vmag^2}{2 \vth^2} - 4\right)\frac{\zeta}{r_B(\Zbar)}
  \frac{1}{\Zbar + 2\zeta}\frac{\mfpe}{L_{T_e}}\fM
  ,
  \label{eq:f1BGK_approximation}
\end{equation}
which is identical to $\xi \ft^1_{\text{Lorentz}}$ for any value of $\zeta$.
One should notice that $r_B(\Zbar\gg1) = \zeta$, i.e. $\zeta$ can be adjusted 
appropriately for example to better address the~transport in nonlocal regime.

\begin{table}
\begin{center}
  \begin{tabular}{c|ccccc}
    \hline\hline\\
    & $\,\Zbar=1\,$ & $\,\Zbar=2\,$ & $\,\Zbar=4\,$ & $\,\Zbar=16\,$ & $\,\Zbar=116\,$ \\\\
    \hline\\
    $\bar{\Delta}\vect{q}_{AWBS}$ & 0.057 & 0.004 & 0.037 & 0.021 & 0.004 \\\\
    \hline\\
    $\phi(\Zbar)$ & -0.037 & -0.003 & 0.04 & 0.058 & 0.065 \\\\ 
    \hline\hline
  \end{tabular}
  \caption{
  Relative error $\bar{\Delta}\vect{q}_{AWBS} = 
  |\vect{q}_{AWBS} - \vect{q}_{SH}| / \vect{q}_{SH}$ of 
  the~$\nue^* = \frac{\nue}{2}$~scaling used in the~AWBS model
  \refeq{eq:AWBS_model} showing the~discrepancy 
  (maximum 6$\%$) with respect to the~original solution of 
  the~heat flux given by numerical solution in Spitzer and Harm 
  \cite{SpitzerHarm_PR1953}. The~values of $\phi(\Zbar)$ (a~weak dependence 
  \eqref{eq:qAWBS_approximation}) are also shown. 
  }
\label{tab:qAWBS}
\end{center}
\end{table}

We have performed an~extensive analysis in the case of 
the~AWBS operator in order to obtain the~heat flux behavior while varying 
$\Zbar$. As expected, the~heat flux magnitude did not match exactly 
the~$\Zbar$-dependence \eqref{eq:qSH_approximation}, e.g. for $\Zbar=1$
the~AWBS heat flux was about 60$\%$ less than the~SH calculation, while
there was a~perfect match in the case of $\Zbar\gg1$. By assuming that the~e-e
collisions are responsible for this inadequacy, we searched for a~scaling of
$\nue$ in \eqref{eq:AWBS_model}. Interestingly, we found an~almost constant
scaling $ r_A$, i.e. with a~very weak dependence on $\Zbar$ as  
\begin{equation}
  \nue^* =  r_A(\Zbar)~\nue 
  = \left(\frac{1}{2} + \phi(\Zbar)\right) \nue \approx \frac{\nue}{2} ,
  \label{eq:qAWBS_approximation}
\end{equation}
where can be approximated as 
$\phi(\Zbar) = \frac{0.59 \Zbar - 1.11}{8.37 \Zbar + 5.15} \ll\frac{1}{2}$ 
for any $\Zbar$, i.e. we decide to use $r_A = \frac{1}{2}$.
Indeed, \tabref{tab:qAWBS} shows $\phi(\Zbar)$ and corresponding relative
error (maximum 6$\%$) of the~heat flux modeled by 
\eqref{eq:AWBS_model} vs. SH results represented by 
\eqref{eq:qSH_approximation}. It should be noted that the~error is calculated 
with respect to original values presented in TABLE~III in 
\cite{SpitzerHarm_PR1953}.  
 
\begin{figure}[tbh]
  \begin{center}
    \begin{tabular}{c}
      \includegraphics[width=0.5\textwidth]{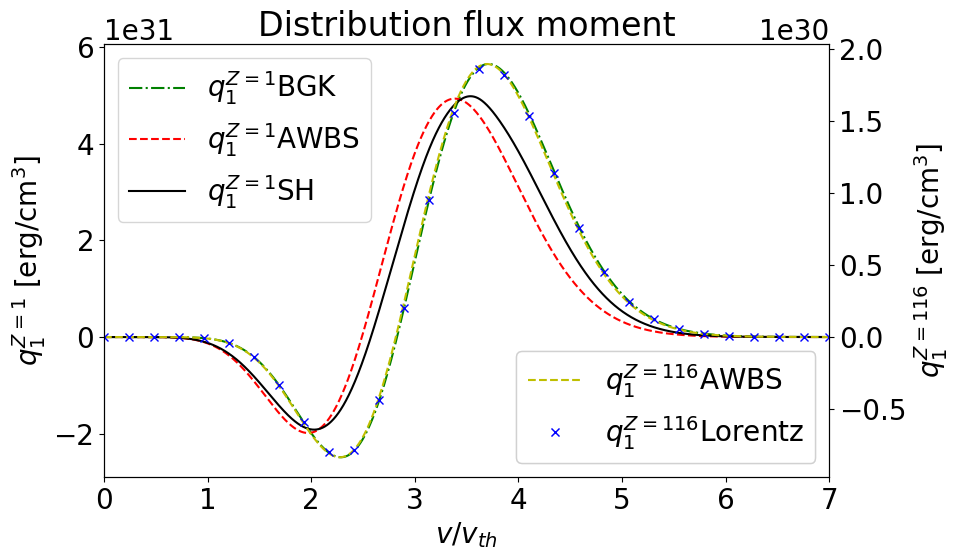}
    \end{tabular}
  \caption{  
  The~flux velocity moment of the~anisotropic part of the~electron distribution 
  function in low $\Zbar=1$ and high $\Zbar=116$ plasmas in diffusive regime. 
  In the~case of $\Zbar = 1$ the~AWBS model matches very well 
  the~reference solution given by the~SH calculation \cite{SpitzerHarm_PR1953} 
  in comparison to the~BGK model. In the~case of $\Zbar = 116$ the~AWBS model
  aligns exactly with the~Lorentz gas approximation as expected. 
  The~BGK and the~SH curves are not shown for $\Zbar = 116$, but also 
  correspond to the~Lorentz gas distribution function.
  }
  \label{fig:q1s_summary}
  \end{center} 
\end{figure}

The~electron-electron collisions scaling 
\cite{Epperlein_PoFB1991} represented by 
\eqref{eq:qSH_approximation} provides only an~integrated information about
the~heat flux magnitude. If one takes a~closer look at the~distribution
function itself, the~conformity of the~modified AWBS collision operator
is even more emphasized as can be seen in \figref{fig:q1s_summary} showing
the~flux moment in function of the~absolute value of velocity
\begin{equation}
  q_1 = \frac{\me\vmag^2}{2}\vmag \ft^1 \vmag^2 .
  \label{eq:q1}
\end{equation}
In the~case of the~high $\Zbar$ plasma ($\Zbar = 116$), 
AWBS exactly aligns with the~Lorentz gas limit \eqref{eq:Lorentz_approximation}.
In the~opposite case of the~low
$\Zbar$ Hydrogen plasma ($\Zbar = 1$), the~AWBS distribution function 
approaches closely the~numerical SH solution \eqref{eq:f1_SH}. 
BGK \eqref{eq:BGK_scaling} takes 
the~Lorentz gas distribution function for any $\Zbar$ 
only scaled by $\xi$.  
The~AWBS collision operator \eqref{eq:AWBS_model} (red dashed line) 
provides 
a~significant improvement with respect to the~SH (Fokker-Planck) solution
\eqref{eq:f1_SH} (solid black line) compared to the~simplest BGK model 
\eqref{eq:f1BGK_approximation} (dashed-dot blue line) 
in \figref{fig:q1s_summary}.

\section{AWBS nonlocal transport model of electrons}
\label{sec:AWBSnonlocal}

In order to define a~nonlocal transport model of electrons, 
we use the~AWBS collision operator and the~P1 angular 
approximation of the~electron distribution function
\begin{equation}
  \tilde{\ft}(\vect{x}, \vn, \vmag) = 
  \fzero(\vect{x}, \vmag) + \vn\cdot\fone(\vect{x}, \vmag) , 
  \label{eq:P1approximation}
\end{equation}
consisting of the~isotropic part represented by the zeroth angular moment 
$\fzero = \frac{1}{4\pi}\int_{4\pi} \tilde{\ft} \dI\vn$ 
and the~directional part represented by the~first angular moment 
$\fone = \frac{3}{4\pi}\int_{4\pi} \vn
\tilde{\ft} \dI\vn$, where $\vn$ is the~transport direction.
Then, the~first two angular moments
\cite{Shkarofsky_Particle_Kinetics_book_1966_24} applied to the~stationary 
form of 
\eqref{eq:kinetic_equation} with collision operator \eqref{eq:AWBS_model} 
(extended by \eqref{eq:qAWBS_approximation}) lead to the~model equations
\begin{eqnarray}
  \vmag\frac{\nue}{2}\pdv{}{\vmag}\left(\fzero - \fM \right) &=&
  \frac{\vmag}{3}\nabla\cdot\fone + \frac{\qe}{\me}\frac{\E}{3}\cdot\left(
  \pdv{\fone}{\vmag} + \frac{2}{\vmag}\fone\right) , 
  \nonumber \\
  \label{eq:AP1f0}\\
  \vmag\frac{\nue}{2}\pdv{\fone}{\vmag}
  - \nuscat\fone &=& 
  \vmag\nabla\fzero + 
  \frac{\qe}{\me}\E\pdv{\fzero}{\vmag} 
  +\frac{\qe\B}{\me c}\vect{\times} \fone
  ,
  \nonumber \\
  \label{eq:AP1f1}
\end{eqnarray}
where $\nuscat = \nuei + \frac{\nue}{2}$. The system of equations 
\eqref{eq:AP1f0} and \eqref{eq:AP1f1} is called the~{\bf AP1 model} 
(AWBS + P1).  

The~AP1 model gives us information about the~electron distribution function 
providing a~bridge between kinetic and fluid description of 
plasma. For example the~\textit{flux} quantities as 
electric current and heat flux due to the~motion of electrons
\begin{equation}
  \vect{j} = \frac{4\pi}{3}\qe \int \vmag \fone \vmag^2\dI\vmag,~~ 
  \vect{q}_h = \frac{4\pi}{3}\frac{\me}{2} \int \vmag^3 \fone \vmag^2\dI\vmag,
  \nonumber
\end{equation}
are based on corresponding velocity moments (integrals) of the~first angular 
moment of EDF. Consequently, the~explicit formula for the~first angular moment
from \eqref{eq:AP1f1} 
proves to be extremely useful
\begin{equation}
  \fone = \frac{\nuscat^2 \vect{F}^* + \omegaB~\omegaB\cdot\vect{F}^* 
  - \nuscat~\omegaB \vect{\times} \vect{F}^*}{\nuscat (\omegaB^2 + \nuscat^2)}
  ,
  \label{eq:f1_explicit}
\end{equation} 
because it provides a~valuable 
information about the~dependence of macroscopic \textit{flux} quantities on
electric and magnetic fields in plasma, 
where $\omegaB = \frac{\qe\B}{\me c}$ is the~electron gyro-frequency and 
$\vect{F}^* = \vmag\frac{\nue}{2}\pdv{\fone}{\vmag} - \vmag\nabla\fzero 
 - \frac{\qe}{\me}\E\pdv{\fzero}{\vmag}$.

\subsection{Nonlocal Ohm's Law}
\label{sec:Efield}
Expression \eqref{eq:f1_explicit} is used
to describe the~electron fluid momentum, i.e. the~current velocity moment
can be written as
\begin{equation}
  \vect{j}_{(f, \E, \B)} = \Iohm\pdv{\fzero}{\vmag}\E 
  + \frac{\me}{\qe} \Iohm\left(\vmag \nabla \fzero 
  - \vmag\frac{\nue}{2}\pdv{\fone}{\vmag} \right)  
  ,
  \label{eq:NonlocalOhm}
\end{equation}
where we used the~following notation 
$\Iohm\vect{g} = - \frac{4\pi \qe^2}{3 \me} \int \vmag \frac{\nuscat^2 \vect{g} 
  + \omegaB~\omegaB\cdot\vect{g} - \nuscat~\omegaB \vect{\times} \vect{g}}
  {\nuscat (\omegaB^2 + \nuscat^2)}~\vmag^2 \dI \vmag$ showing 
how the~operator $\Iohm$
acts on a~general vector field $\vect{g}$.
We refer to 
\eqref{eq:NonlocalOhm} as to the~{\bf nonlocal Ohm's law}.
The~need for a~nonlocal Ohm's law to accurately capture magnetic field 
advection due to the~Nernst effect has been demonstrated 
\cite{Luciani85, Ridgers08, Brodrick18}. A~full investigation 
of this new Ohm's law is beyond the~scope of this article. 
The~high $\Zbar$ ($\nue \ll \nuei$) local asymptotic to 
the~standard Ohm's law  
can be found when $\fzero\rightarrow\fM$ and weak magnetization 
($\omegaB \ll \nuei$) is considered. Then \eqref{eq:NonlocalOhm} simplifies to
\begin{multline}
  \vect{j} =- \frac{\qe^2}{\me} \int \frac{\vmag^3}{\nuei}
  \left( \E~\pdv{\fM}{\vmag} + \frac{\me}{\qe}\vmag \nabla \fM \right)
  ~\dI \vmag =\\
  \frac{16\sqrt{\frac{2}{\pi}}\qe^2 \kB^\frac{3}{2} \Te^\frac{3}{2}}{\me^\frac{5}{2} \Gamma\Zbar}
  \left[\E - \frac{\frac{5}{2} \ed \kB \nabla \Te 
  + \nabla \ed \kB \Te}{\qe \ed}  \right] 
  ,
  \label{eq:AsymptoticOhm}
\end{multline}
which can be directly compared to the~local fluid theory  
\begin{equation}
  \E = \matr{\sigma}(\fzero)^{-1} \vect{j} 
  - \frac{\nabla P (\fzero)}{\qe \ed}
  \xrightarrow{\fzero \rightarrow \fM}
  \E_{l} = \frac{\vect{j}}{\sigma_{l}}
  + \frac{\nabla p_e - \vect{R}_{\Te}}{\qe \ed} 
  ,
  \label{eq:GeneralOhm} 
\end{equation}
where the~local electric field $\E_{l}$ is given by
the~pressure $p_e = \ed \kB \Te$,
the~thermal force $\vect{R}_{\Te} = - \frac{3}{2} \ed \kB \nabla \Te$ 
and the~local electrical conductivity 
$\sigma_{l} = 16\sqrt{\frac{2}{\pi}}\qe^2 \kB^\frac{3}{2} \Te^\frac{3}{2}
/ \me^\frac{5}{2} \Gamma\Zbar$ \cite{Braginskii_1965_3}.
In \eqref{eq:GeneralOhm} we defined the~nonlocal electrical tensor conductivity 
\begin{equation}
  \matr{\sigma} = \Iohm\pdv{\fzero}{\vmag}
  ,
  \label{eq:NonlocalSigma}
\end{equation}
and the~nonlocal microscopic force
\begin{equation}
  \nabla P = \matr{\sigma}^{-1}\me\ed \Iohm\vmag \nabla \fzero
  ,
  \label{eq:NonlocalGradP}
\end{equation}
based on \eqref{eq:NonlocalOhm}.

The~local dependence of the~AP1 current \eqref{eq:AsymptoticOhm} 
on electric field and gradients of $\ed$ and $\Te$ clearly demonstrates, 
that \eqref{eq:GeneralOhm} is a~local version of \eqref{eq:NonlocalOhm}.
This also implies that \eqref{eq:NonlocalOhm} provides 
a~magnetic field source in terms of nonlocal Biermann battery,
since the~curl on the~electric field 
\eqref{eq:GeneralOhm} gives
\begin{equation}
  \nabla\vect{\times} \frac{\nabla P}{\qe \ed} 
  \xrightarrow{\fzero \rightarrow \fM} 
  \nabla\vect{\times} \frac{\nabla p_e - \vect{R}_{\Te}}{\qe \ed} =
  \frac{\kB}{\qe\ed}\nabla \Te \vect{\times}\nabla \ed
  .
  \label{eq:NonlocalBiermann}
\end{equation}
The~nonlocal Biermann battery effect \eqref{eq:NonlocalBiermann} 
can lead to a~spontaneous magnetic field generation under uniform 
density plasma profile as has been shown in \cite{Kingham_PRL2002}.

A~local version of the~{\bf nonlocal Ohm's law} \eqref{eq:NonlocalOhm} 
compared to the~\textit{generalized Ohm's law} \eqref{eq:GeneralOhm} with
a~magnetic field is deferred to a~future complementary work.

\subsection{AWBS Nonlocal Magneto-Hydrodynamics}
\label{sec:ANTH}
The~\textit{AWBS nonlocal~magneto-hydrodynamic~model} (Nonlocal-MHD)
refers to two~temperature single-fluid hydrodynamic model 
extended by a~kinetic model of electrons using the~AWBS transport equation,
which provides a~direct coupling between hydrodynamics and Maxwell equations.

Mass, momentum density, and total energy 
$\rho$, $\rho\vect{u}$, and 
$E = \frac{1}{2}\rho\vect{u}\cdot\vect{u} + 
 \rho \varepsilon_i + \rho \varepsilon_e$, 
where $\rho$ is 
the~density of plasma, $\vect{u}$ the~plasma fluid velocity, $\varepsilon_i$ 
the~specific internal ion energy density, 
and $\varepsilon_e$ the~specific internal 
electron energy density,
are modeled by the~Euler equations in the~Lagrangian frame 
\cite{Holec_DGBGKT_2016, Holec_PoPNTH2018}
\begin{eqnarray}
 \frac{\dI \rho}{\dI t} &=& - \rho\nabla\cdot\vect{u}
 , 
 \label{eq:NTH_rho}\\
 \rho\, \frac{\dI \vect{u}}{\dI t} &=& - \nabla (p_i + p_e) 
 + \vect{j}_{(f, \E, \B)} \vect{\times}\B
 ,  
 \label{eq:NTH_v}\\
 \rho~C_{V_i}\frac{\dI \Ti}{\dI t} 
 &=& 
 \left(\rho^2 C_{\Ti} - p_i\right)\nabla\cdot\vect{u} 
 - G(\Ti - \Te)
 ,  
 \label{eq:NTH_Ti}\\
 \rho~C_{V_e} \frac{\dI \Te}{\dI t} 
  &=& 
 \left(\rho^2 C_{\Te} - p_e \right) \nabla\cdot\vect{u}  
 + G(T_i - \Te)
 \nonumber\\ 
 && - \nabla\cdot\vect{q}_{h(f, \E, \B)} + Q_{\text{IB}} 
 , 
 \label{eq:NTH_Te}
\end{eqnarray}
where $\Ti$ is the~temperature of ions, $\Te$ the~temperature of electrons,
$p_i$ the~ion pressure, $p_e$ the~electron pressure,
$\vect{q}_h$ the~heat flux, $Q_{\text{IB}}$ the~inverse-bremsstrahlung laser 
absorption (which can also distort the distribution function away from 
a~Maxwellian \cite{Langdon80}, strongly modifying the~transport 
\cite{Ridgers08_2}, an~effect which will not be considered further here) and 
$G = \rho C_{V_e} \nuei$ is 
the~ion-electron energy exchange rate. 
The~thermodynamic closure terms 
$p_e$, $p_i$, 
$C_{V_i} = \frac{\partial \varepsilon_i}{\partial \Ti}$, 
$C_{\Ti} = \pdv{\varepsilon_i}{\rho}$,
$C_{V_e} = \frac{\partial \varepsilon_e}{\partial \Te}$, 
$C_{\Te} = \pdv{\varepsilon_e}{\rho}$
are obtained from an~equation of state (EOS), e.g.
the~SESAME equation of state tables
\cite{T4_SESAME_83, Lyon_SESAME_EOS_database-TechRep-92}.

The~magnetic and electric fields are modeled by Maxwell equations
\begin{eqnarray}
  \frac{1}{c}\frac{\partial \B}{\partial t} + \nabla\vect{\times}\E &=& 0 
  ,
  \label{eq:Faraday} \\
  \nabla\vect{\times}\B - \frac{4\pi}{c} \vect{j}_{(f, \E, \B)} &=& 0
  .
  \label{eq:Ampere}
\end{eqnarray}

We have explicitly written the~current and heat flux as dependent on
electron kinetics, represented by the~electron distribution function $f$,
and electric and magnetic fields. In principal, $\vect{j}_{(f, \E, \B)}$
and $\vect{q}_{h(f, \E, \B)}$ can be referred to as the~\textit{kinetic closure}
and is provided by the~{\bf AP1 model} \eqref{eq:AP1f0} and \eqref{eq:AP1f1}.

\subsection{Numerical Implementation of the AWBS Electron Kinetics}
\label{sec:Numerics}

Proceeding further, one can make use of 
the~{\bf nonlocal Ohm's law} \eqref{eq:NonlocalOhm} to write 
a~\textit{fully kinetic form of Ampere's law} 
governing the~electric field $\E$
\begin{equation}
  \Iohm\pdv{\fzero}{\vmag}\E 
  + \frac{\me}{\qe} \Iohm\vmag \nabla \fzero = 
  \frac{c}{4\pi} \nabla\vect{\times}\B 
  .
  \label{eq:AmpereKinetic}
\end{equation}

In order to solve the~kinetics of electrons, we adopt a~high-order 
finite element discretization 
\cite{Dobrev_Kolev_Rieben-High-order_curvilinear_finite_element_methods_for_Lagrangian_hydrodynamics, mfem-library} 
of the~model equations \eqref{eq:AP1f0}, \eqref{eq:AP1f1}, 
\eqref{eq:AmpereKinetic}
\begin{eqnarray}
  \matr{M}^{L_2}_{_{(\frac{\vmag\nue}{2})}}\cdot\ddv{\vfzero}{\vmag} 
  - \matr{V}^{L_2}_{_{(\frac{\qe\E}{3\me})}}\cdot\ddv{\fone}{\vmag}
  &=&
  \matr{D}^{L_2}_{_{(\frac{\vmag}{3})}}\cdot\fone 
  + \matr{M}^{L_2}_{_{(\frac{2\qe\E}{3\me\vmag})}}\cdot\fone
  \nonumber \\ 
  &&+ \vect{b}^{L_2}_{_{(\frac{\vmag\nue}{2}\pdv{\fM}{\vmag})}}, 
  \label{eq:FEMAP1f0}
  \\
  \matr{M}^{H_1}_{_{(\frac{\vmag\nue}{2})}}\cdot\ddv{\fone}{\vmag}
  - \matr{V}^{H_1}_{_{(\frac{\qe\E}{\me})}}\cdot\ddv{\vfzero}{\vmag}
   &=& 
  \matr{G}^{H_1}_{_{(\vmag)}}\cdot\vfzero 
  + \matr{M}^{H_1}_{_{(\nuscat)}}\cdot\fone 
  \nonumber \\
  && + \matr{C}^{H_1}_{_{(\frac{\qe\B}{\me c}\vect{\times})}}\cdot\fone
  ,
  \label{eq:FEMAP1f1}\\
  \matr{J}^{ND}_{_{(\pdv{\fzero}{\vmag})}}\cdot\E 
  &=& 
  \matr{JG}^{ND}_{_{(\frac{\me\vmag}{\qe})}}\cdot\vfzero
  + \vect{b}^{ND}_{_{(\frac{c}{4\pi} \nabla\vect{\times}\B)}} 
  ,
  \nonumber \\
  \label{eq:FEMAmpereKinetic}
\end{eqnarray}
where the~continuous differential operators are represented by standard 
discrete analogs (matrices of bilinear forms) 
$\matr{M}, \matr{G}, \matr{D}, \matr{V}, \matr{C}$, i.e. mass, gradient, 
divergence, vector field dot product, and vector field curl, and
by $\matr{J}, \matr{JG}$ matrices specific to {\bf nonlocal Ohm's law} 
\eqref{eq:NonlocalOhm}. The~linear form $\vect{b}$ represents sources, i.e.
temperature $\Te$ via $\pdv{\fM}{\vmag}$ and the~curl of 
the~magnetic field $\B$. These finite element discrete analogs are defined
on piece-wise continuous $L_2$ finite element space (domain of $\vfzero$),
continuous $H_1$ finite element space (domain of $\fone$) 
\cite{Dobrev_Kolev_Rieben-High-order_curvilinear_finite_element_methods_for_Lagrangian_hydrodynamics}, 
and Nedelec finite element space (domain of $\E$). We do not show their
definitions since it is out of the~scope of this article. 

The~strategy of solving 
\eqref{eq:FEMAP1f0} and \eqref{eq:FEMAP1f1} resides in integrating 
$\ddv{\vfzero}{\vmag}$
and $\ddv{\fone}{\vmag}$ along the~velocity axis. 
This is done by starting the~integration
from the~maximum  velocity 
($\vmag = 7 \vth^{max}$ is a~sufficiently high limit) 
to zero velocity using the~Implicit Runge-Kutta method. The~value
$\vth^{max}$ equals the~electron thermal velocity corresponding to the~maximum 
electron temperature in the~current profile of plasma.
It should be noted, that the~backward integration concept is crucial for 
the~model, since it corresponds to the~deceleration of electrons due to 
collisions \cite{Touati_2014}. Consequently, we refer to 
\textit{decelerating} AP1 model, which however, leads to the~limitation of 
the~electric field described in \appref{app:AP1limit}. 

\section{Benchmarking the~AWBS nonlocal transport model}
\label{sec:BenchmarkingAWBS}
Having shown several encouraging properties of the~AWBS transport 
equation defined by \eqref{eq:AWBS_model} under local diffusive conditions
in \secref{sec:DiffusiveKinetics}, this section focuses on analyzing 
its behavior under nonlocal plasma conditions, extensively investigated 
in numerous publications 
\cite{Malone_1975_15, Colombant_PoP2005, Bell_1981_83, LMV_1983_7, Brantov_Nonlocal_electron_transport_1998, Schurtz_2000, Sorbo_2015}.
A~variety of tests suitable for benchmarking the~nonlocal electron 
transport models have been published 
\cite{Epperlein_PoFB1991, marocchino2013, Sorbo_2015, 
Sorbo_2016, Sherlock_PoP2017, Brodrick_PoP2017}, we focus on 
conditions relevant to inertial confinement fusion plasmas generated by lasers.

We show results of our implementation of the~AP1 nonlocal transport model 
presented in \secref{sec:AWBSnonlocal} benchmarked against simulation results 
provided by a~rather complete set of kinetic models with varying complexity. 
The~most reliable models represents a~collisional Particle-In-Cell
code Calder \cite{Lefebvre_NF2003, Perez_PoP2012} resolving 
the~plasma frequency time scale, and a~standard VFP codes
Aladin and Impact \cite{Kingham_JCP2004}.
In addition, we compare the~SNB nonlocal transport model \cite{Schurtz_2000} 
used in hydrodynamic codes. 
That is a~first time when a~collisional PIC code is used
for benchmarking of nonlocal electron transport models. 

\textbf{Calder PIC code}

The~particle evolution in the~phase-space, including small angle 
binary collisions, is described with 
the~Maxwell equations \eqref{eq:Faraday}, \eqref{eq:Ampere}
coupled with the~ion and electron Vlasov equations with 
the~Landau-Beliaev-Budker collisions integral (LBB)
\cite{Landau_1936, Beliaev_SPD1956} 
\begin{multline}
\frac{\partial f_\alpha}{\partial t}+\mathbf{v}\cdot\nabla_{\mathbf{x}}f_\alpha+q_\alpha\left(\mathbf{E}+\mathbf{v}\vect{\times}\mathbf{B}\right)\nabla_{\mathbf{p}}f_\alpha =
\\
C_{LBB}(f_\alpha,f_\alpha)+\sum_\beta C_{LBB}(f_\alpha,f_\beta)
.
\end{multline}
The LBB collision integral takes the~form
\begin{multline}
C_{LBB}(f_\alpha,f_\beta)=
\\
-\frac{\partial}{\partial \mathbf{p}}\cdot\frac{\Gamma_{\alpha\beta}}{2}\left[\int \mathbf{U}(\mathbf{p},\mathbf{p}^\prime)\cdot(f_\alpha\nabla_{\mathbf{p}^\prime}f_\beta^\prime-f_\beta^\prime\nabla_{\mathbf{p}}f_\alpha)\right]d^3\mathbf{p}^\prime
,
\label{eq:LBB_model}
\end{multline}
where its relativistic kernel reads
$\mathbf{U}(\mathbf{p},\mathbf{p}^\prime)=\frac{r^2/\gamma\gamma^\prime}{(r^2-1)^{3/2}}$ 
$\left[(r^2-1)\mathbf{I}-\mathbf{p}\otimes\mathbf{p}-\mathbf{p}^\prime\otimes\mathbf{p}^\prime+r(\mathbf{p}\otimes\mathbf{p}^\prime+\mathbf{p}^\prime\otimes\mathbf{p})\right]$
with $\gamma=\sqrt{1+\mathbf{p}^2}$, $\gamma^\prime=\sqrt{1+\mathbf{p}^{\prime 2}}$ and $r=\gamma\gamma^\prime-\mathbf{p}\cdot\mathbf{p}^\prime$. 
The momemtum $\mathbf{p}_\alpha$ ($\mathbf{p}_\beta$) is normalized to 
$m_\alpha c$ (resp. $m_\beta c$). The~collision operator \eqref{eq:LBB_model} 
tends to \eqref{eq:LFP_model} in the non-relativistic limit.
The aforementioned model is solved in 3D by the PIC code CALDER. 
\cite{Lefebvre_NF2003, Perez_PoP2012}.

\textbf{Impact and Aladin VFP codes}

PIC simulations are extremely expensive as the~collisions require description 
of the~velocity space in 3 dimensions. Yet, a~reduction of dimensions can be 
done by developing the~distribution function in a~Cartesian tensor series, 
equivalent to expansion in the spherical harmonics \cite{Johnston_PR1960}.
The~first order form corresponds to the~P1 approximation 
\eqref{eq:P1approximation} and coupled with
the~Landau-Fokker-Planck collisional operator 
\eqref{eq:LFP_model} leads to the P$_1$-VFP model 
\cite{Johnston_PR1960, Kingham_JCP2004}:
\begin{eqnarray}
\frac{\partial \fzero}{\partial t}+\frac{\vmag}{3}\nabla \cdot \fone
+\frac{\qe}{3\me\vmag^2}\frac{\partial}{\partial \vmag}(\vmag^2\E\cdot \fone)
&=&
C^0_{ee}(\fzero) ,
 \label{eq:P1f0_Aladin}\\
\frac{\partial \fone}{\partial t}+\vmag\nabla \fzero
+ \frac{\qe\E}{\me}\frac{\partial \fzero}{\partial \vmag}
+\frac{\qe\B}{\me}\vect{\times} \fone 
&=&
- \nuei\fone .
\label{eq:P1f1_Aladin}
\end{eqnarray}
where only the~isotropic part of the~distribution function in 
the~e-e collision integral \eqref{eq:LFP_model} is used
\begin{eqnarray} 
C^0_{ee}(\fzero) &=& \frac{\Gamma}{v^2}\frac{\partial}{\partial \vmag}
\left[C(f_0)f_0+D(\fzero)\frac{\partial \fzero}{\partial \vmag}\right] ,
\label{eq:C0_collision_operator}
\\
C(\fzero(\vmag)) &=& 4\pi\int_0^\vmag \fzero(u) u^2 \dI u ,
\nonumber
\\
D(\fzero(\vmag)) &=& \frac{4\pi}{\vmag}\int_0^\vmag u^2\int_u^\infty w \fzero(w) 
\dI w \dI u .
\nonumber
\end{eqnarray}

The~codes Impact and Aladin solve the system \eqref{eq:P1f0_Aladin} 
and \eqref{eq:P1f1_Aladin} with the Maxwell equations  
\eqref{eq:Faraday} and \eqref{eq:Ampere} in two spatial dimensions, 
assuming immobile ions.

The~model AP1 uses similar equations as Aladin
and Impact with the~difference, that AP1 describes the~steady-state 
electron distribution function with respect to the~ions,
and is using a~simplified (linear) collision operator inherently
coupled to ions via the~hydrodynamic equations.


\textbf{SNB approach}

Now considered as a~standard nonlocal electron transport models in hydrodynamic 
codes, SNB \cite{Schurtz_2000} represents an efficient P1 method based on
the~velocity dependent form of the~collision BGK operator. It uses EDF 
approximation representing deviation from the~local BGK theory
\begin{equation}
  \tilde{\ft} = 
  \fM + \delta\fzero 
  + \vn\cdot\left(\fone_M + \delta\fone\right) . 
  \label{eq:SNB_approximation}
\end{equation}
Equations for the~zero and first angular moments follow from 
the~electron transport equation with scaled collision operator 
\eqref{eq:BGK_scaling}
according to the~SNB approximation \eqref{eq:SNB_approximation} 
(similar to \eqref{eq:AP1f0} and \eqref{eq:AP1f1})
\begin{eqnarray}
  r_B \delta\fzero &=&
  - \frac{\vmag}{3}\nabla\cdot\delta\fone
  - \frac{\vmag}{3}\nabla\cdot\fone_M
  \nonumber\\ 
  &&\xcancel{- \frac{\qe}{\me}\frac{\E}{3}\cdot\left(
  \pdv{\fone_M}{\vmag} + \pdv{\delta\fone}{\vmag} 
  + \frac{2}{\vmag}\left(\fone_M + \delta\fone\right)\right)} , 
  \nonumber \\
  \label{eq:SNBf0}\\
  \frac{\nuei}{\xi}\delta\fone &=& - \vmag\nabla\delta\fzero 
  ~\xcancel{- \frac{\qe}{\me}\E\pdv{\delta\fzero}{\vmag}}
  \nonumber \\
  &&\underbrace{- \frac{\nuei}{\xi}\fone_M - \vmag\nabla\fM
  - \frac{\qe}{\me}\E\pdv{\fM}{\vmag}
  }_{~~~~~~~~~~~~~=~0~defines~\fone_M} 
  ,
  \nonumber \\
  \label{eq:SNBf1}
\end{eqnarray}
where the~magnetic field was neglected, the~under-braced part
of \eqref{eq:SNBf1} when $\delta\fzero$ and $\delta \fone$ are zero
defines the~local anisotropic term
\begin{equation}
  \fone_M = -\xi\mfpei\fM\left( \frac{\nabla\ed}{\ed} 
+ \left( \frac{\vmag^2}{2\vth^2} - \frac{3}{2}\right)
\frac{\nabla\Te}{\Te} - \frac{\qe\E}{\me\vth^2}\right) ,
  \label{eq:f1M}
\end{equation}
and the~efficiency of SNB resides in omitting the~electric field 
effect (crossed out terms in \eqref{eq:SNBf0} and \eqref{eq:SNBf1}), which
leads to a~simple diffusion equation for the~correction to the~isotropic 
part of the~distribution function
\begin{equation}
  \frac{1}{\mfpe^{SNB}}\delta\fzero 
  - \nabla\cdot\frac{\mfpei^{SNB}}{3}\nabla\delta\fzero =
  \nabla\cdot\frac{\xi\mfpei}{3}\fM\frac{\nabla \Te}{\Te}
  ,
  \label{eq:SNB_model}
\end{equation}
where $\frac{1}{\mfpei^{SNB}} = 
\frac{\nuei}{\xi\vmag} + \frac{|\qe\E|}{\frac{1}{2}\me\vmag^2}$ 
and $\mfpe^{SNB} = \frac{\vmag}{r_B \nue}$, and the~source term based on 
$\fone_M$ simplifies by avoiding the~electric field effect, density gradient 
and the~$\vmag$-dependent bracket in \eqref{eq:f1M}. The~missing
effect of $\E$ in \eqref{eq:SNB_model} is accounted for by 
an~isotropic scattering in definition of $\mfpei^{SNB}$ \cite{Schurtz_2000}.
Consequently, the~effect of the~electric field in SNB
is accounted for only via $\fone_M$, where the~electric field is
fixed to $\E_L$. 

As shown previously, the~BGK collision operator \eqref{eq:BGK_scaling}
provides one free parameter $\zeta$.
We propose to use $\zeta = 2$ giving $r_B(\Zbar\gg1) = 2$ which agrees 
with $r=2$ in SNB formulation proposed in \cite{Brodrick_PoP2017} 
for the~case of ICF relevant plasma. We also have $r_B(\Zbar = 1) = 1.677$,
which means that our pure kinetic derivation of SNB varies just slightly
from a~constant value $r_B=2$ \cite{Brodrick_PoP2017}, yet it provides 
slightly better results. The~explicit form of the~anisotropic 
part of EDF then reads $\fone = \fone_M - \mfpei^{SNB}\nabla\delta\fzero$.

\begin{figure*}[htb]
  \begin{center}
    \begin{tabular}{cc}
        \includegraphics[width=\figscale\textwidth]{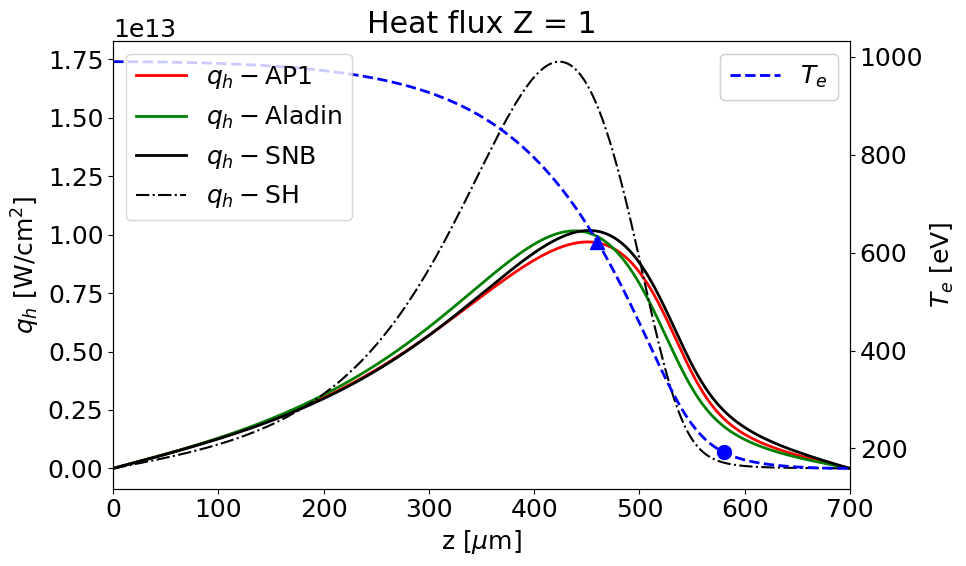} 
	  &
      \includegraphics[width=\figscale\textwidth]{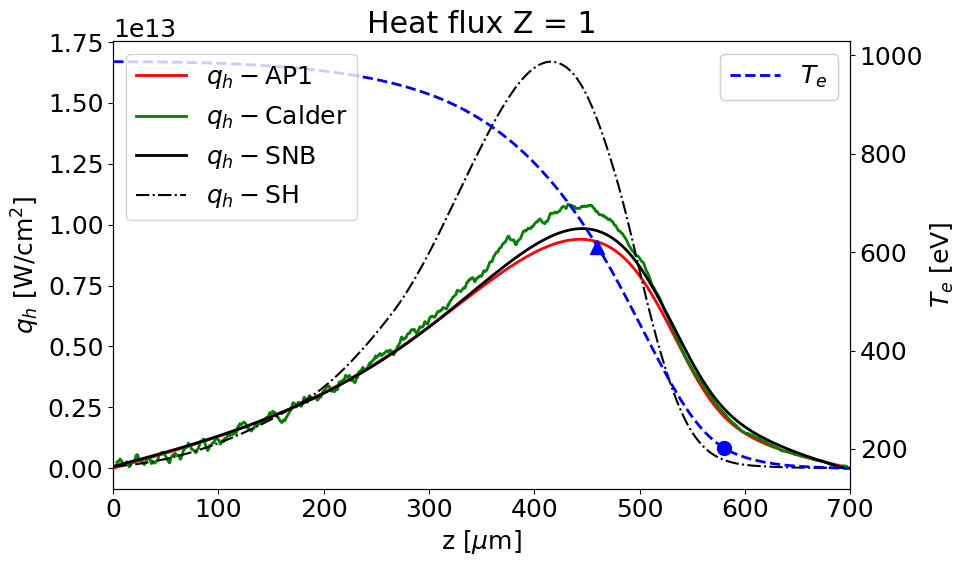} 
	  \\ 
      \includegraphics[width=\figscale\textwidth]{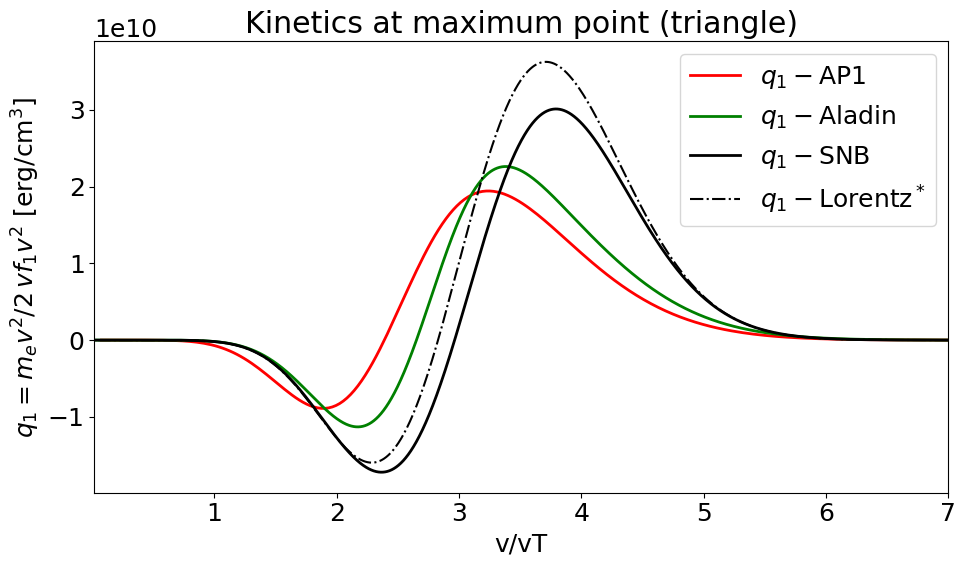} 
	  &
      \includegraphics[width=\figscale\textwidth]{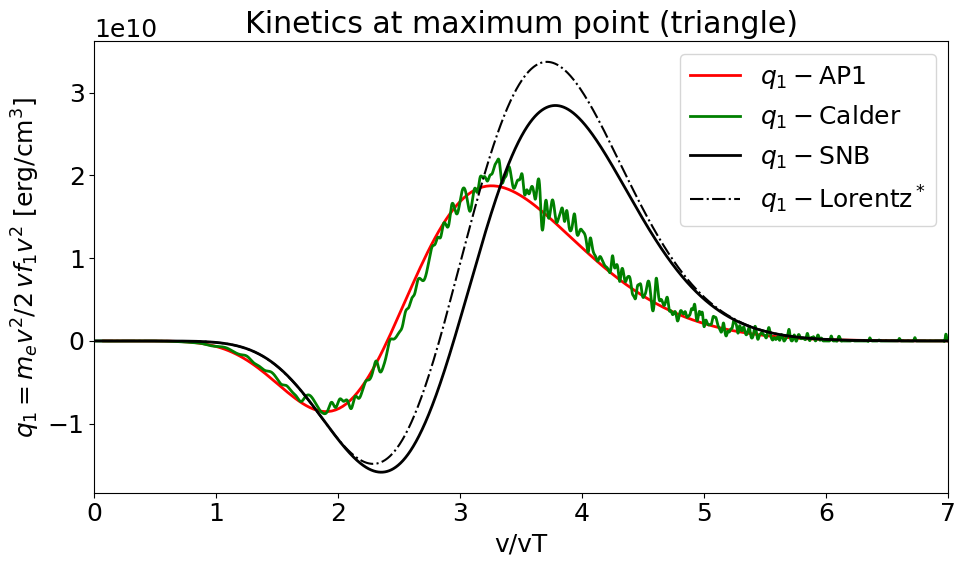} 
	  \\
 \includegraphics[width=\figscale\textwidth]{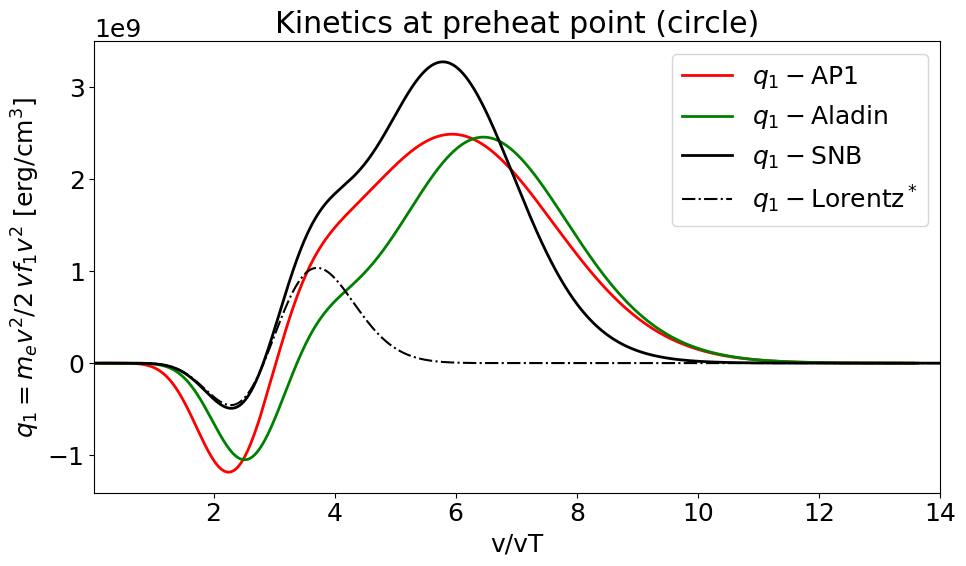} 
      &  \includegraphics[width=\figscale\textwidth]{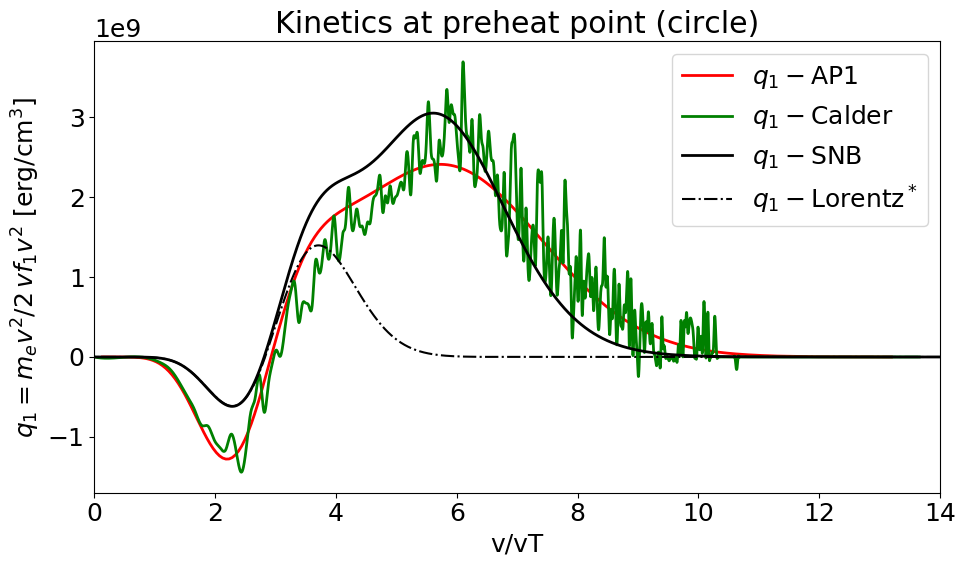}
	\end{tabular}
  \caption{  
  AP1 performance in a~low-$\Zbar$ heat-bath problem compared to the~VFP code 
  Aladin (left) and the~collisionl PIC code Calder (right). 
  The~heat flux and temperature
  profiles at 20 ps are shown in the~top plots also for AP1 and SNB.
  Middle and bottom plots show a~kinetic detail of the~anisotropic part
  of EDF (its flux velocity moment) at two different spatial points. 
  The~results of the~local Lorentz gas theory scaled by the~SH correction
  are also shown for reference. An~excellent agreement in EDF between
  AP1 (AWBS collision operator \eqref{eq:AWBS_model}) 
  and Calder (full Landau-Fokker-Planck collision operator 
  \eqref{eq:LFP_model}) is observed.
  }
  \label{fig:C7_AladinCalder_case5}
  \end{center} 
\end{figure*}

\subsection{Heat-bath problem}  
\label{sec:heatbath_test}
AP1 is compared to Calder, Aladin, Impact, and SNB by 
calculating the~heat flow in the~case of a~homogeneous plasma
with a~large temperature variation
\begin{equation}
  T_e(z) = 0.575 - 0.425 \tanh\left((z-450) s\right) ,
  \label{eq:T_init}
\end{equation}
which exhibits a~steep gradient at the~point 450~$\mu$m 
connecting a~hot bath ($T_e = 1$~keV) 
and cold bath ($T_e = 0.17$~keV) and $s$ is the~parameter of steepness. 
This test is referred to as a~simple non-linear heat-bath problem and
originally was introduced in \cite{marocchino2013} and further investigated
in  \cite{Sorbo_2015, Sorbo_2016, Sherlock_PoP2017, Brodrick_PoP2017}.
\begin{figure}[htb]
  \begin{center}
    \begin{tabular}{c}
      \includegraphics[width=\figscale\textwidth]{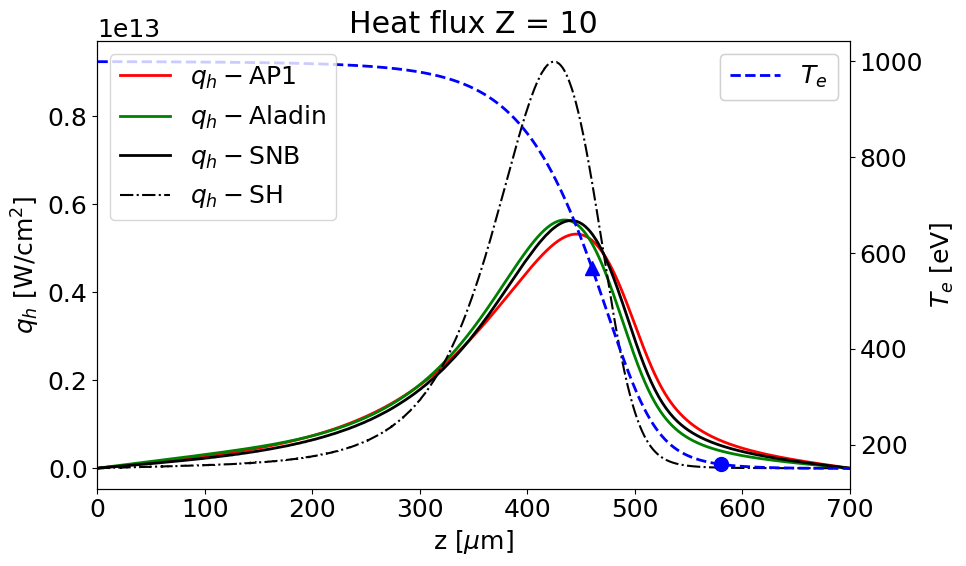} \\
      \includegraphics[width=\figscale\textwidth]{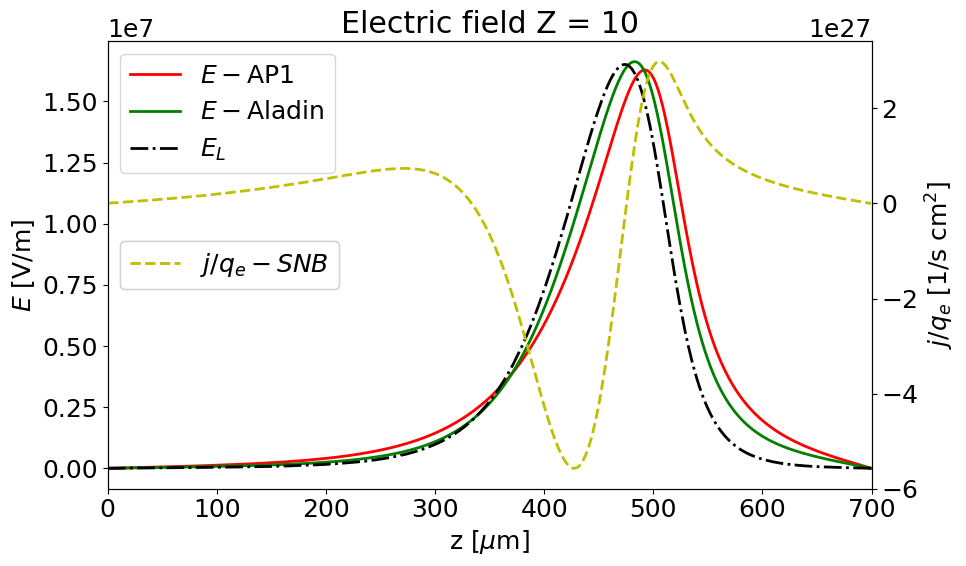} \\
      \includegraphics[width=\figscale\textwidth]{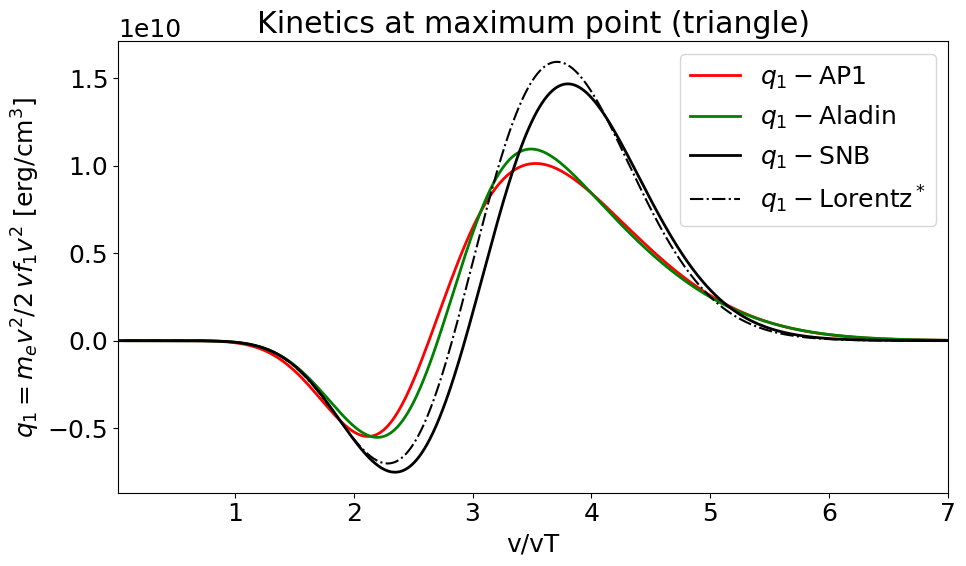} \\
      \includegraphics[width=\figscale\textwidth]{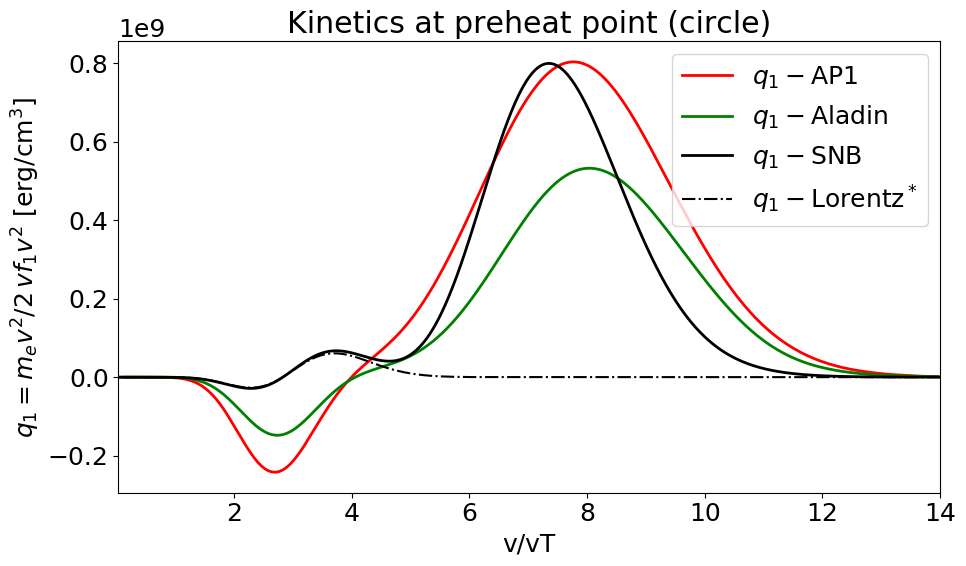}  
    \end{tabular}
  \caption{  
  A~moderate-$\Zbar$ heat-bath problem. The~temperature profile evolved 
  up to 12 ps by Aladin.
  Top plots show heat flux profiles and electric fields by AP1, Aladin, 
  and SNB. The~resulting current of SNB using explicitly the~local electric 
  field $\E_L$ is also shown. Botom plots show a~kinetic detail of
  the~anisotropic part of EDF (its flux velocity moment) at two different 
  spatial points by AP1 and SNB compared to Aladin. 
  }
  \label{fig:C7_Aladin_case3}
  \end{center} 
\end{figure}

The~total computational box size is 700~$\mu$m.
We performed Aladin, Impact, and Calder simulations showing an~evolution of
temperature starting from the~initial profile \eqref{eq:T_init}. 
Due to the~initial distribution function being approximated by a~Maxwellian,
the~first phase of the~simulation exhibits a~transient behavior of the~heat
flux. After several ps the~distribution adjusts to its asymptotic form
and the~heat flux profiles can be compared. 
We then take the temperature profiles from Aladin/Impact/Calder and compare 
with AP1 and SNB models which calculate a~stationary heat flow
for a~given temperature profile. 
For all heat-bath simulations the electron density, Coulomb logarithm and 
ionisation were kept constant and uniform.
The~Coulomb logarithm was held fixed throughout, $\lnc = 7.09$.

We show AP1 results for two ionization states, namely $\Zbar = 1$ and 
$\Zbar = 10$ 
in \figref{fig:C7_AladinCalder_case5} and \figref{fig:C7_Aladin_case3}, 
respectively, corresponding to a~moderate nonlocality 
(Kn$^e \sim 10^{-2}$) leading to a~roughly 40 $\%$ inhibition compared 
to the~local SH heat flux maximum. 
A~constant $\ed = 5\times10^{20}$~cm$^{-3}$ is held throughout the~simulation 
and the~original temperature profile steepness $s = 1/50~\mu$m.
It is preferable to use 
$\text{Kn}^e = \frac{\mfpe(\vth)}{\sqrt{\Zbar + 1}L_{T_e}}$ instead of
 $\text{Kn} = \frac{\mfpei(\vth)}{L_{T_e}}$, 
because $\sqrt{\Zbar + 1}$ provides 
a~better scaling of nonlocality with respect
to ionization \cite{LMV_1983_7}, i.e. the~flux inhibition and Kn$^e$ are
kept approximately the~same when varying $\Zbar$ in 
\figref{fig:C7_AladinCalder_case5} and \figref{fig:C7_Aladin_case3}.
In addition to the~heat flux profiles, we also show the~distribution function 
details related to the~approximate point of the~heat flux maximum (460 $\mu$m) 
and to the~point of the~nonlocal preheat effect (580 $\mu$m) in the~form of
the~flux moment of EDFs anisotropic part \eqref{eq:q1}.
The~nonlocal preheat effect shows a~very good agreement with 
previous results published in \cite{Sherlock_PoP2017}.

The~top left plot of \figref{fig:C7_AladinCalder_case5} shows heat flux 
profiles computed by Aladin, AP1, and SNB corresponding to the~temperature 
$\Te$ profile computed by Aladin and the~top right plot of 
\figref{fig:C7_AladinCalder_case5} shows heat flux profiles computed by Calder, 
AP1, and SNB corresponding to the~temperature $\Te$ profile computed by Calder.
Both kinetic simulations by Aladin and Calder evolved up to 20 ps for 
$\Zbar = 1$. 
The~anisotropic part of EDF, in particular, 
the~heat flux velocity moment $q_1$, 
at the~heat flux maximum (triangle point) and at~the~nonlocal preheat region 
(circle point) computed by AP1 and SNB for the~temperature profiles by 
Aladin and Calder, can be used as a~detailed comparison of four conceptually
different models: the~full anisotropy form \eqref{eq:LFP_model} of 
the~FP collision operator (Calder); the~isotropic form 
\eqref{eq:C0_collision_operator} of the~FP collision operator (Aladin);
the~simplified linear form \eqref{eq:AWBS_model} of the~FP collision operator
(AWBS in AP1); 
the~nonlocal electron transport model \eqref{eq:SNB_model} (SNB). 
Excellent match of $q_1$ can be seen between 
AP1 and Calder at the~both spatial points. On the~other hand, 
the~AP1 profiles of EDF provide a~reasonable match to Aladin too, however,
one observes a~deviation
which resembles to the~low $\Zbar$ trend shown in \figref{fig:q1s_summary}, 
where AP1 corresponds to AWBS and Aladin to BGK curves. To summarize,
various FP-like codes are compared in detail, in particular collisional PIC 
for the first time, and all show a~very good match. 
Furthermore, the~effect of the~anisotropy in the~collision model, 
captured by AP1 and neglected by Aladin and Impact, proves to be important
in the~low-$\Zbar$ plasma. 

In the~case $\Zbar = 10$, we show heat flux profiles
computed by Aladin, AP1, and SNB corresponding to the~temperature $\Te$
profile computed by Aladin up to 12 ps 
in the~top plot of \figref{fig:C7_Aladin_case3}. 
Corresponding profiles of 
a~self-consistently calculated electric fields by Aladin and AP1 
(using the~nonlocal Ohm's law) are shown in the~higher middle plot.
Also the~local theory based electric field $\E_L$ used by SNB is shown.
EDF at the~point of the~approximate heat flux maximum (triangle) of
the~temperature profile is shown in the~lower middle plot, 
where a~very precise match between AP1 and Aladin can be observed, 
and $q_1$ at the~preheat point (circle) of the~temperature profile is shown 
in the~bottom plot. In the~latter case AP1 shows a~very similar properties 
as Aladin with a~difference in magnitude corresponding to a~higher 
heat flux computed by AP1 at this point. 

SNB shows very good results of the~heat flux profile in all three cases, 
i.e. compared to Aladin and Calder in \figref{fig:C7_AladinCalder_case5} 
and to Aladin in \figref{fig:C7_Aladin_case3}. 
However, one can observe that the~EDF kinetic solution of SNB provides 
only a~qualitative image with respect to the~reference green line solution. 
This is illustrated for example in
\figref{fig:C7_Aladin_case3}, where the~kinetics at preheat point plot 
reveals an~insufficient electric field treatment (no return current). 
The~kinetics at maximum point plot shows that the~solution SNB solution
approaches closely the~local Lorentz$^*$ solution and that 
significantly recedes from the~reference fully kinetic solution (green line).
These~discrepancies can be attributed to the~use of an~inconsistent 
electric field in the~case of SNB which uses $\E_L$. An~electric field 
comparison is shown in \figref{fig:C7_Aladin_case3}, where it is shown that 
the~local electric field treatment $\E_L$ used in SNB fails 
in the~preheat region and consequently leads to 
a~significant violation of the~plasma quasi-neutrality,
i.e. a~non-zero current, where one can observe an~uncontrolled stream of 
electrons in the~preheat and also an~overestimation of negative return current
around the~heat flux maximum.

The~AP1 model equations \eqref{eq:AP1f0}, \eqref{eq:AP1f1}, 
and \eqref{eq:AmpereKinetic} in general show a~very good performance 
in all three cases when compared to the~fully kinetic results 
(green line) by Aladin and Calder, which can be assigned 
to the~AWBS collision operator and the~consistent 
treatment of $\E$ via nonlocal Ohm's law \eqref{eq:NonlocalOhm} in
 \eqref{eq:AmpereKinetic} (no $\B$ field in 1D).

\begin{figure}[htb]
  \begin{center}
    \begin{tabular}{c}
      \includegraphics[width=\figscale\textwidth]{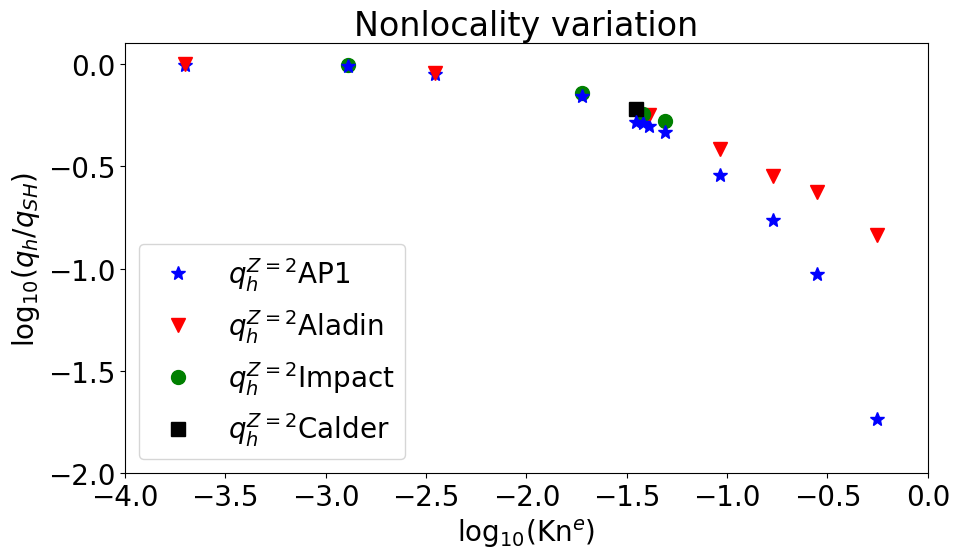}
    \end{tabular}
  \caption{  
  The~heat-flux inhibition compared to the~local SH theory along varying
  nonlocality (Kn$^e$) in the~heat bath problem. AP1 compares well to 
  the~full kinetic simulations by Aladin, Impact, and Calder, up to 
  Kn$^e \sim 10^{-1}$. For higher nonlocality \textit{decelerating} 
  AP1 departs significanlty from the~reference solutions because of
  the~electric field limiting in accordance with the~velocity limit in 
  \tabref{tab:vlim}. 
  }
  \label{fig:Kn_results}
  \end{center} 
\end{figure}

In addition, the~Knudsen number Kn$^e$ has been varied among the~simulation 
runs in order to address a~broad range of nonlocality of 
the~electron transport corresponding 
to the~laser-heated plasma conditions, i.e. Kn$^e \in (0.0001, 1)$. 
The~variation of Kn$^e$ arises from the~variation
of the~uniform electron density $n_e \in (10^{19}, 10^{23})$~cm$^{-3}$ or 
the~length scale given by the~slope of the~temperature profile 
$s \in (1/2500, 1/25)~\mu$m. Results showing the~heat flux maximum 
of an~extensive set of simulations of
varying Kn$^e$ is shown in \figref{fig:Kn_results}.
When analyzing the~simulation results shown in \figref{fig:Kn_results}, 
we observed that 
the~maximum of $q_1$ at the~maximum point tends to decrease with increasing 
Kn$^e$ and that the~interval of electron velocities important for 
the~heat transport always belongs to $3 \vth < \vmag <4 \vth$ for an~example
refer to the~kinetics at maximum point in \figref{fig:C7_AladinCalder_case5} 
and \figref{fig:C7_Aladin_case3}. 
According to simulations,
the~stopping force in \eqref{eq:AP1f0} and \eqref{eq:AP1f1} is dominated by
the~electric field for electrons with velocity above the~velocity limit
\begin{equation}
  \vmag_{lim} = \sqrt{\frac{\sqrt{3}\Gamma\me}{2\qe}\frac{n_e}{|\E|}}
  ,
  \label{eq:v_limit}
\end{equation}
and this limit drops 
down significantly with increasing Knudsen number as can be seen 
in \tabref{tab:vlim}. 
As a~consequence, the~electrons responsible for the~heat flux
($3 \vth < \vmag <4 \vth$) are preferably affected by the~electric field
rather than by collisions when Kn$^{e} > 10^{-1}$. According to 
\tabref{tab:vlim} collsions dominate stopping for $\vmag < 3.1 \vth$ 
when Kn$^e = 10^{-1}$ and even a~much lower value $\vmag < 1.8 \vth$ 
when Kn$^e = 1.0$. This explains the~unsatisfactory results of 
the~\textit{decelerating} AP1
model for high Kn$^e$ shown in \figref{fig:Kn_results}. 
Notably, the~AP1 limited electric field effect (described in 
\appref{app:AP1limit}) leads to a~steep increase of error with respect 
to VFP code Aladin for Kn$^e > 10^{-1}$. 
For example $\vmag_{lim}\sim 4.3\vth$ for the~maximum point EDF in 
\figref{fig:C7_AladinCalder_case5}.

Unfortunately, \eqref{eq:v_limit} also leads to a~limitation of 
the~\textit{decelerating} AP1 model, where the~strength of the~stopping/accelerating effect due to the~electric field on electrons must always 
be kept less than the~e-e collision friction. Details are shown 
in \appref{app:AP1limit}.

\begin{table}
\begin{center}
  \begin{tabular}{c|ccccc}
    \hline\hline\\
    Kn$^e$ & $\,\,10^{-4}\,\,$ & $\,\,10^{-3}\,\,$ & $\,\,10^{-2}\,\,$ & $\,\,10^{-1}\,\,$ & $\,\,1\,\,$ \\\\
    \hline\\
    $\vmag_{lim} / \vth$ & 70.8 & 22.4 & 7.3 & 3.1 & 1.8\\\\
    \hline\hline
  \end{tabular}
  \caption{
  Scan over varying nonlocality (Kn$^e$) showing the~limit of 
  the~collision friction dominance over the~deceleration of electrons 
  due to the~electric field force. The~electric field effect is dominant
  for electrons with higher velocity than $\vmag_{lim}$ defined in 
  \eqref{eq:v_limit}. Kn$^e$ and $\vth$ are evaluated from the~same 
  plasma profiles.
  }
\label{tab:vlim}
\end{center}
\end{table}

\subsection{Hohlraum problem}
Additionally to the~steep temperature gradients, the~laser-heated plasma 
experiments also involve steep density gradients and variation in ionization,
which are dominant effects in multi-material hohlraums
at the interface between the helium gas-fill and 
the ablated high $\Zbar$ plasma.

\begin{figure}[htb]
  \begin{center}
    \begin{tabular}{c}
      \includegraphics[width=\figscale\textwidth]{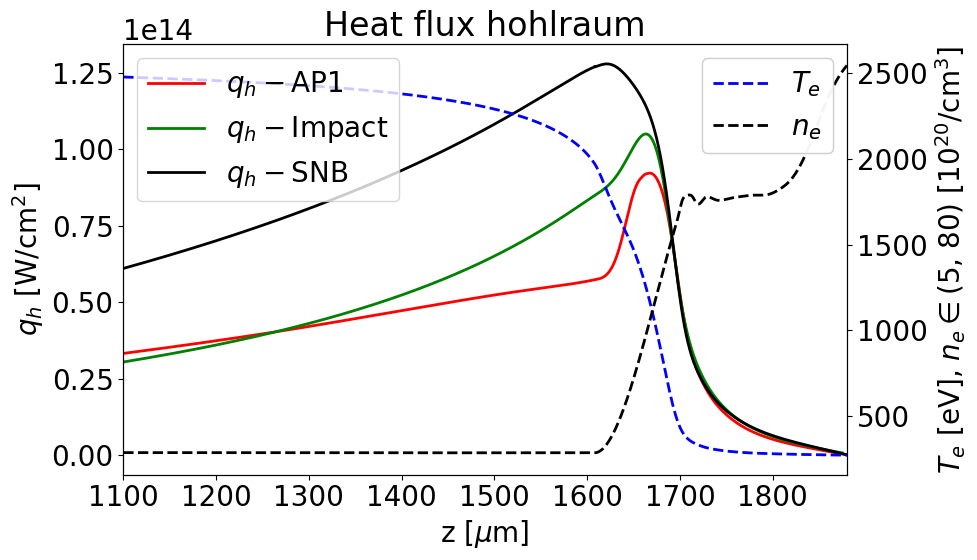} 
    \end{tabular}
  \caption{
  Heat flux profiles by AP1, Impact and SNB along 
  the~electron temperature $\Te$ and electron density $\ed$
  profiles in a~laser-heated gadolinium hohlraum 
  with a~helium gas-fill.
  }
  \label{fig:Gd_VFP_10ps_heatflux}
  \end{center} 
\end{figure}

In~\cite{Brodrick_PoP2017}, a~kinetic simulation of laser pulse interaction 
with a gas filled hohlraum was presented. 
Plasma profiles provided by a~HYDRA simulation in 1D
geometry of a~laser-heated gadolinium hohlraum containing a~helium 
gas at time of 20 ns were used as input for the~Impact 
\cite{Kingham_JCP2004} VFP code.  
For simplicity, the Coulomb logarithm was treated as a
constant $\lnc_{ei}$ = $\lnc_{ee}$ = 2.1484. In reality, in the~low-density 
corona $\lnc$ reaches 8, which, however, does not affect the~heat flux profile 
significantly. 
\figref{fig:Gd_VFP_10ps_heatflux} shows the~electron temperature $\Te$ 
evolved during 10 ps by Impact and the~electron density $\ed$ profile.
Along with plasma profiles the~heat flux profiles
of AP1, Impact, and SNB are also shown.  

One can observe a~very good
match between AP1 and Impact computations in the~preheat region.
It is worth mentioning that in the~surroundings of the~heat flux maximum 
($\sim 1662~\mu$m) the~profiles of all plasma variables 
exhibit steep gradients 
with a change from $T_e$ = 2.5 keV, $n_e$ = 5$\times$10$^{20}$ cm$^{−3}$, 
$\Zbar$ = 2 to $T_e$ = 0.3 keV, $n_e$ = 6$\times$10$^{21}$ cm$^{−3}$ , 
$\Zbar$ = 44 across approximately 100 $\mu$m 
(between 1600~$\mu$m and 1700~$\mu$m), starting at the~helium-gadolinium 
interface.  
In this region, we can see a~qualitative match between AP1 and Impact 
providing a~same sign of the~heat flux divergence, however,
the~electric field limitation explained in 
\appref{app:AP1limit} leads to a~stronger drop of 
the~\textit{decelerating} AP1 heat flux on the~material 
interface, which then closely aligns to the~Impact heat flux in the~corona. 
On the~other hand, 
SNB overestimates significantly the~heat flux in the~lower density part 
of plasma up to the~point of the~heat flux maximum given by Impact 
(green line in \figref{fig:Gd_VFP_10ps_heatflux}). More 
importantly, SNB shows the~opposite sign of the~heat flux divergence compared to Impact
(and AP1) in the~steep gradients region close to the~material interface. 
In the~preheat region SNB performs 
very well. Nevertheless, it is important to stress that 
SNB required only 25 velocity groups compared to 250 velocity groups used by
Impact and AP1 for this ICF relevant plasma conditions, thus making it a~very
efficient modeling approach though its description of kinetics is rather
qualitative.

\section{Conclusions}
\label{sec:Conclusions}
In conclusion, we have performed a~thorough analysis of the~AWBS transport 
equation for electrons originally introduced in \cite{Sorbo_2015} and extended
it by adding a~nonlocal version of Ohm's law.
After redefining the~e-e collission term, we have shown that the~AWBS
simplified linear form of the~Fokker-Planck collision operator keeps
important kinetic properties in local diffusive regime. 
It provides a~correct dependence 
on the~ion charge $\Zbar$ (BGK requires an~additional fix) 
and inherently includes
the~anisotropic part of the~distribution function $\fone$, which compares
very well to the~full Fokker-Planck operator.
Under nonlocal transport plasma conditions, we benchmarked AP1 against 
the~reference VFP codes Aladin and Impact, collisional PIC code Calder, 
and the~standard nonlocal approach SNB. This is a~first time quantitative
comparison of collisional PIC and VFP codes. 
AP1 performed very well over all simulation cases while capturing 
the~important kinetic features compared to the~reference kinetic codes. 
Furthermore, our detailed analysis of the~anisotropic part of the~EDF 
provided by AP1 showed an~excellent match with Calder and outperformed
Aladin and Impact in the~case of low-$\Zbar$ plasma, which is attributed
to the~effect of anisotropy in the~collision model. 
This suggests 
a~promising AP1's capability in predicting general transport coefficients and 
the~seeding of parametric laser plasma instabilities sensitive 
to the~Landau damping of longitudinal plasma waves 
\cite{goldston1995introduction, Sorbo_2015},
which is of great importance in ICF related plasmas 
\cite{Kirkwood_NIFLPI_PPCF2013}.
Other kinetic effects as perpendicular transport, e.g heat flow
or magnetic field advection, occurring in magnetised plasma  
\cite{Walsh_Nernst_PRL2017} are introduced in
AP1 via the~nonlocal Ohm's law, which recovers 
the~generelized Ohm's law in the~local diffusive asymptotic limit.
The~importance of the~nonlocal Ohm's law becomes obvious for Kn$^e > 10^{-1}$,
where the~stopping of nonlocal electrons is rather due to the~electric field 
effect than the~collisional friction.
We have also shown a~new formulation of SNB based on
the~scaled BGK collision operator \eqref{eq:BGK_scaling}, which performed
well in the~heat-bath problem and the~corresponding heat flux profile. 
However, EDF output is rather qualitative which also lead to 
non-precise results of the~hohlraum problem.
We also observed an~inaccurate kinetic results of 
the~\textit{decelerating} AP1 computation for highly nonlocal plasma 
conditions, which is explained by the~velocity limit applied to the~action
of the~electric field.

\begin{acknowledgments}
This work was performed under the auspices of the U.S. Department of Energy by Lawrence Livermore National Laboratory under Contract DE-AC52-07NA27344.
This work was partially supported by the project ELITAS (ELI Tools for Advanced Simulation) CZ.02.1.01/0.0/0.0/16$\_$013/0001793 from the European Regional Development Fund.
C.~P.~Ridgers would like to acknowledge funding from EPSRC (grant EP/M011372/1). This work has been carried out within the framework of the EUROfusion Consortium and has received funding from the Euratom research and training programme 2014–2018 under grant agreement No 633053 (project reference CfP-AWP17-IFE-CCFE-01). The views and opinions expressed herein do not necessarily reflect those of the European Commission.

This document was prepared as an account of work sponsored by an agency of the United States government. Neither the United States government nor Lawrence Livermore National Security, LLC, nor any of their employees makes any warranty, expressed or implied, or assumes any legal liability or responsibility for the accuracy, completeness, or usefulness of any information, apparatus, product, or process disclosed, or represents that its use would not infringe privately owned rights. Reference herein to any specific commercial product, process, or service by trade name, trademark, manufacturer, or otherwise does not necessarily constitute or imply its endorsement, recommendation, or favoring by the United States government or Lawrence Livermore National Security, LLC. The views and opinions of authors expressed herein do not necessarily state or reflect those of the United States government or Lawrence Livermore National Security, LLC, and shall not be used for advertising or product endorsement purposes.
\end{acknowledgments}

\appendix
\section{Analysis of local diffusive regime}
\label{app:DiffusiveKinetics}

In order to analyze the~local diffusive regime, 
we use the~BGK collision operator 
\eqref{eq:BGK_model_1D} 
\begin{equation}
  \frac{1}{\vmag}C(\tilde{\ft})
  =
  \frac{\fM - \tilde{\ft}}{\mfpe}
  + \frac{1}{2}\left(\frac{\Zbar}{\mfpe} + \frac{1}{\mfpe}\right)
  \pdv{}{\mu}(1 - \mu^2)\pdv{\tilde{\ft}}{\mu} 
  ,\nonumber
\end{equation}
to write explicitly \eqref{eq:1D_kinetic_equation} 
for \eqref{eq:f_approximation} 
\begin{multline}
  \frac{\qe\Ez}{\me\vmag^2} \ft^1 
  + \mu^2\left[\pdv{\ft^1}{z} 
  + \frac{\qe\Ez}{\me\vmag}\pdv{\ft^1}{\vmag} - \frac{\qe\Ez}{\me\vmag^2} \ft^1
  \right] \\ 
  + \mu\left[\pdv{\ft^0}{z} 
  + \frac{\qe\Ez}{\me\vmag}\pdv{\ft^0}{\vmag}\right] = 
  \frac{\fM - \ft^0}{\mfpe} - \mu \frac{\Zbar + 2}{\mfpe}\ft^1
  .
  \label{app_eq:BGK_model_1D}
\end{multline}
The~P1 form \eqref{eq:f_approximation} represents a~low anisotropy
expansion to the~first to Legendre polynomials $P_0 = 1$ and $P_1 = \mu$, 
where the~projection of a~function $f(\mu)$ 
to a~Legendre polynomial $P_k(\mu)$ reads
$\mathcal{P}_k(f) = \int_{-1}^1 P_k(\mu) f(\mu) \dI\mu$, in particular giving
the~orthogonality $\mathcal{P}_0(P_1) = \mathcal{P}_1(P_0) = 0$.

Consequently, the~projections of the~equation \eqref{app_eq:BGK_model_1D}, 
i.e. $\mathcal{P}_0\eqref{app_eq:BGK_model_1D}$ 
and $\mathcal{P}_1\eqref{app_eq:BGK_model_1D}$, define 
\begin{eqnarray}
    \ft^0 &=& \fM - \frac{\mfpe}{3}\left[\frac{2\qe\Ez}{\me\vmag^2} \ft^1 
  + \pdv{\ft^1}{z} + \frac{\qe\Ez}{\me\vmag}\pdv{\ft^1}{\vmag}\right]
  ,
  \label{app_eq:BGK_f0} \\
  \ft^1 &=& - \frac{\mfpe}{\Zbar + 2}
  \left[ \pdv{\ft^0}{z} + \frac{\qe\Ez}{\me\vmag}\pdv{\ft^0}{\vmag} \right]
  . 
  \label{app_eq:BGK_f1}
\end{eqnarray}
It is valid to assume that $\ft^0 \approx \fM$, i.e. that $\fM \gg
\frac{\mfpe}{3}\left[\pdv{\ft^1}{z} +
\frac{\qe\Ez}{\me\vmag^3}\pdv{\vmag^2\ft^1}{\vmag}\right]$ 
in \eqref{app_eq:BGK_f0}. The~\textit{quasi-neutrality} constraint 
\eqref{eq:j0_P1} applied to \eqref{app_eq:BGK_f1} along with 
$\ft^0 = \fM$ leads to the~electric field 
(same as the~classical Lorentz electric field $\E_L$ \cite{Lorentz_1905})
\begin{equation}
  \Ez = \frac{\me\vth^2}{\qe}\left(\frac{1}{L_{n_e}} 
  + \frac{5}{2}\frac{1}{L_{T_e}} \right) 
  ,
  \label{app_eq:BGK_Efield}
\end{equation}
and the~anisotropic part of EDF takes the~form \eqref{eq:BGK_approximation}.
It should be noticed that $\ft^0$ equilibrates to $\fM$ 
as $O\left( \text{Kn}^2\right)$ since $\ft^1 \sim \text{Kn} \fM$ 
and $\mfpe \Ez \sim \text{Kn}$.

The~AWBS operator \eqref{eq:AWBS_model} applied to 
\eqref{eq:f_approximation} reads
\begin{eqnarray}
  \frac{1}{\vmag}C_{AWBS}(\tilde{\ft})
  &=& 
  \frac{\vmag r_A}{\mfpe} \pdv{}{\vmag}\left(\tilde{\ft} - \fM\right) 
  \nonumber \\
  && + \frac{1}{2}\left(\frac{\Zbar}{\mfpe} + \frac{r_A}{\mfpe}\right)
  \pdv{}{\mu}(1 - \mu^2)\pdv{\tilde{\ft}}{\mu}  \nonumber \\
  &=& \frac{\vmag r_A}{\mfpe} \pdv{}{\vmag}\left(\ft^0 - \fM\right) \nonumber \\ 
  &&\, + \mu\left(\frac{\vmag r_A}{\mfpe} \pdv{\ft^1}{\vmag} 
  - \frac{\Zbar+r_A}{\mfpe}\ft^1\right) ,
  \label{app_eq:AWBS_model_1D}
\end{eqnarray}
where $\nue^* = r_A \nue = \frac{\vmag r_A}{\mfpe}$ with $r_A$ being a~scaling
parameter of the~standard e-e collision frequency. 
The~$\mathcal{P}_0$ and $\mathcal{P}_1$ projections 
of the~equation \eqref{app_eq:BGK_model_1D} using 
\eqref{app_eq:AWBS_model_1D} instead of BGK then define
\begin{eqnarray}
  \pdv{}{\vmag}\left( \ft^0 -\fM\right) &=& 
  \frac{\mfpe}{3}\left[\pdv{\ft^1}{z} +
  \frac{\qe\Ez}{\me\vmag^3}\pdv{\vmag^2\ft^1}{\vmag}\right] ,
  \label{app_eq:AWBS_f0} \\
  \pdv{\ft^1}{\vmag} 
  - \frac{\Zbar + r_A}{{\vmag r_A}}\ft^1 &=& \frac{\mfpe}{\vmag r_A}
  \left[\pdv{\ft^0}{z} + \frac{\qe\Ez}{\me\vmag}\pdv{\ft^0}{\vmag}\right] 
  .
  \label{app_eq:AWBS_f1} 
\end{eqnarray}
If we assume that $\pdv{\ft^0}{\vmag} = \pdv{\fM}{\vmag}$, i.e. $\ft^0 = \fM$,
the~anisotropic part of the~AWBS operator is governed by the~equation
\eqref{eq:AWBS_f1}, which can be simplified to the~form
\begin{equation}
  \pdv{\vmag^a \ft^1}{\vmag} = 
  \frac{\mfpe \vmag^{a-1}}{r_A}
  \left( b~\frac{\vmag^2}{2 \vth^2} + c\right)\fM ,
  \nonumber 
\end{equation}
with an~integral solution (using $\ft^1(\infty) = 0$)
\begin{equation}
  \ft^1(v) = - \frac{d}{\vmag^a} 
  \int_{\frac{\vmag^2}{2\vth^2}}^\infty (b \tilde{v}^{\frac{a+6}{2} - 1} 
  + c \tilde{v}^{\frac{a+4}{2} - 1}) \exp(-\tilde{v}) d\tilde{v} 
  ,
  \label{app_eq:gammaInt}
\end{equation}
where the~analytical solution to \eqref{app_eq:gammaInt} can be obtained 
in the~form of upper incomplete gamma function shown 
in~\secref{sec:AWBSDiffusiveRegime}, where the~coefficients 
$a, b, c,$ and $d$ are defined. 
However, since the~analytical formula \eqref{eq:AWBS_analytic_solution} 
is valid for $a>-4$, we also adopt the~implicit Euler 
numerical integration with $\Delta v < 0$, where we integrate from high 
electron velocity ($v_{max} = 7\vth$) to zero  
(using 10$^5$ $\Delta v$ steps). The~numerical approach is used for the~case
of $\Zbar > 1.5$ and $r_A = \frac{1}{2}$.

\section{AP1 electric field limit}
\label{app:AP1limit}

We have encountered a~very specific property of the~AP1 model
with respect to the~electric field magnitude. The~easiest way how to 
demonstrate this is to write the~model equations \eqref{eq:AP1f0} and 
\eqref{eq:AP1f1} in 1D (z-axis). Then, due to its linear nature, it is easy 
to eliminate one of the~partial derivatives with respect to $\vmag$, i.e. 
$\pdv{\fzero}{\vmag}$ or $\pdv{\fonez}{\vmag}$. 
In the~case of elimination of $\pdv{\fzero}{\vmag}$ 
one obtains the~following equation
\begin{multline}
  \left(\vmag\frac{\nue}{2} - \frac{2\qe^2\Ez^2}{3\me^2\vmag\nue}\right) 
  \pdv{\fonez}{\vmag} 
  =
  \frac{2\qe\Ez}{3\me\nue}\pdv{\fonez}{z}  
  + \frac{4\pi\qe\Ez}{3\me}\pdv{\fM}{\vmag} \\
  + \frac{\vmag}{3}\pdv{\fzero}{z} 
  + \left(\frac{4\qe^2\Ez^2}{3\me^2\vmag^2\nue}
  + \left(\nuei + \frac{\nue}{2}\right) \right)\fonez .
  \label{eq:AP1_model_1D}
\end{multline}
It is convenient to write the~bracket on the~left hand side of 
\eqref{eq:AP1_model_1D} as
$\frac{2}{3\vmag\nue} 
\left(\left(\sqrt{3}\vmag\frac{\nue}{2}\right)^2 
- \frac{\qe^2}{\me^2}\Ez^2\right)$
from where it is clear that the~bracket is negative if 
$\sqrt{3}\vmag\frac{\nue}{2} < \frac{\qe}{\me}|\E|$, 
i.e. there is a~velocity limit for a~given magnitude $|\E|$, 
when the~collisions are no more fully dominant and the~electric field 
introduces a~comparable effect to the~collision friction in 
the~electron transport.

It can be shown, that the~last term on the~right hand side of 
\eqref{eq:AP1_model_1D} is dominant and the~solution behaves as 
\begin{equation}
  \Delta \fone \sim \exp\left(\frac{\frac{4\qe^2\Ez^2}{3\me^2\vmag^2\nue}
  + \left(\nuei + \frac{\nue}{2}\right)}
  {\vmag\frac{\nue}{2} - \frac{2\qe^2\Ez^2}{3\me^2\vmag\nue}}\, 
  \Delta\vmag\right) ,
  \label{eq:f1z_behavior}
\end{equation}
where $\Delta \vmag < 0$ represents a~velocity step of the~implicit Euler
numerical integration of decelerating electrons.
However, \eqref{eq:f1z_behavior} exhibits an~exponential growth 
for velocities above the~friction limit (bracket on the~left hand side of 
\eqref{eq:AP1_model_1D})
\begin{equation}
  \vmag_{lim}  = \sqrt{\frac{\sqrt{3}\Gamma\me}{2\qe}\frac{n_e}{|\E|}} ,
  \label{app_eq:v_limit}
\end{equation}
which makes the~problem to be ill-posed.

In order to provide a~stable model, we introduce a~reduced electric field
to be acting as the~accelerating force of electrons
\begin{equation}
  |\E_{red}| = \sqrt{3} \vmag\frac{\me}{\qe}\frac{\nue}{2} ,
  \label{eq:Elimit}
\end{equation}
ensuring that the~bracket on the~left hand side of \eqref{eq:AP1_model_1D}
remains positive. We define a~quantity $\eta_{red} = \frac{|\E_{red}|}{|\E|}$.
Then, the~AP1 model \eqref{eq:AP1f0}, \eqref{eq:AP1f1} can be formulated 
as well posed 
\begin{eqnarray}
  \vmag\frac{\nue}{2}\pdv{}{\vmag}\left(\fzero - \fM \right) &=&
  \frac{\vmag}{3}\nabla\cdot\fone + \frac{\qe}{\me}\frac{\E}{3}\cdot
  \nonumber \\
  &&\left(
  \eta_{red}\pdv{\fone}{\vmag} + \frac{2(2-\eta_{red})}{\vmag}\fone\right) , 
  \nonumber \\
  \label{eq:AP1f0_app}\\
  \vmag\frac{\nue}{2}\pdv{\fone}{\vmag}
  - \nuscat\fone &=& 
  \vmag\nabla\fzero + 
  \frac{\qe\eta_{red}}{\me}\E\pdv{\fzero}{\vmag} 
  +\frac{\qe\B}{\me c}\vect{\times} \fone
  ,
  \nonumber \\
  \label{eq:AP1f1_app}
\end{eqnarray}
while introducing the~reduction factor of the~accelerating electric field
and the~compensation of the~electric field effect via its angular term.

\bibliographystyle{elsarticle-num}
\bibliography{NTH}

\clearpage

\end{document}